\documentclass[superscriptaddress, aps, pra
, longbibliography,twocolumn, footinbib]{revtex4-2}
\usepackage{amssymb}
\usepackage{amsmath}

\usepackage{upgreek}
\usepackage{color}
\usepackage{graphicx}
\usepackage{float}
\usepackage{nccmath}
\usepackage{bm}
\usepackage{hyperref}
\usepackage[mathlines]{lineno}
\usepackage{dcolumn}
\usepackage[normalem]{ulem}
\usepackage{braket}

\hypersetup{colorlinks,breaklinks,
            urlcolor=[rgb]{0,0,0.36},
            linkcolor=[rgb]{0,0,0.36},
            citecolor=[rgb]{0,0,0.36}}

\makeatother

\begin{document}

\title{Efficient variational approach to the Fermi polaron problem in two
dimensions,\\
both in and out of equilibrium}
\author{Yi-Fan Qu}
\affiliation{CAS Key Laboratory of Theoretical Physics, Institute of
Theoretical Physics, Chinese Academy of Sciences, Beijing 100190, China} %
\affiliation{School of Physical Sciences, University of Chinese Academy of
Sciences, Beijing 100049, China}
\affiliation{Institute for Theoretical Physics, ETH Zurich, 8093 Zurich,
Switzerland}
\author{Pavel E. Dolgirev}
\affiliation{Department of Physics, Harvard University, Cambridge,
Massachusetts 02138, USA}
\affiliation{Institute for Theoretical Physics, ETH Zurich, 8093 Zurich,
Switzerland}
\author{Eugene Demler}
\affiliation{Institute for Theoretical Physics, ETH Zurich, 8093 Zurich,
Switzerland}
\author{Tao Shi}
\email{tshi@itp.ac.cn}
\affiliation{CAS Key Laboratory of Theoretical Physics, Institute of
Theoretical Physics, Chinese Academy of Sciences, Beijing 100190, China} %
\affiliation{CAS Center for Excellence in Topological Quantum Computation,
University of Chinese Academy of Sciences, Beijing 100049, China}

\date{\today }

\begin{abstract}
We develop a non-Gaussian variational approach that enables us to study both
equilibrium and far-from-equilibrium physics of the two-dimensional Fermi
polaron. This method provides an unbiased analysis of the
polaron-to-molecule phase transition without relying on truncations in the
total number of particle-hole excitations. Our results -- which include the
ground state energy and quasiparticle residue -- are in qualitative
agreement with the known Monte Carlo calculations. The main advantage of the
non-Gaussian states compared to conventional numerical methods is that they
enable us to explore long-time polaron evolution and, in particular, study
various spectral properties accessible to both solid-state and ultracold
atom experiments. We design two types of radiofrequency spectroscopies to
measure polaronic and molecular spectral functions. Depending on the
parameter regime, we find that these spectral functions and fermionic
density profiles near the impurity display either long-lived oscillations
between the repulsive and attractive polaron branches or exhibit fast
relaxational dynamics to the molecular state.
\end{abstract}

\maketitle

\affiliation{CAS Key Laboratory of Theoretical Physics, Institute of
Theoretical Physics, Chinese Academy of Sciences, Beijing 100190, China}
\affiliation{School of Physical Sciences, University of Chinese Academy of
Sciences, Beijing 100049, China}

\affiliation{Department of Physics, Harvard University, Cambridge,
Massachusetts 02138, USA}

\affiliation{Institute for Theoretical Physics, ETH Zurich, 8093 Zurich,
Switzerland}

\affiliation{CAS Key Laboratory of Theoretical Physics, Institute of
Theoretical Physics, Chinese Academy of Sciences, Beijing 100190, China}
\affiliation{CAS Center for Excellence in Topological Quantum Computation,
University of Chinese Academy of Sciences, Beijing 100049, China}

\section{Introduction}

Fermi polaron models correspond to a general class of quantum many-body
problems in which a single impurity interacts with a bath of fermions.
Historically, theoretical work in this area started with the analysis of
models with infinitely heavy impurities, which exhibit a phenomenon of
orthogonality catastrophe~\cite{anderson1967}. The latter is an observation of P. W. Anderson
that even a weak impurity potential results in the creation of an infinite
number of low-energy particle-hole excitations. Orthogonality catastrophe
plays an important role in several areas of physics, including X-ray
scattering \cite{ohtaka1990,mahan2000}, photoemission \cite%
{Anderson1969,Anderson1970,tanabe1985orthogonality}, transport in mesoscopic systems \cite%
{hentschel2005,nazarov2009quantum}, and radiofrequency (RF) and Rydberg spectroscopies in
ultracold atoms \cite{schirotzek2009,zhang2012,Richard2016,Yuto2019}.
Dynamics in polaronic systems becomes even richer when impurity particles
are endowed with internal degrees of freedom. The simplest example is adding
spin states to a localized impurity, corresponding to the Kondo model. This
class of systems exhibits such striking phenomena as non-monotonic
temperature dependence of resistivity in metals with magnetic impurities
\cite{sarachik1964}, formation of heavy fermion materials \cite%
{hewson1993kondo}, and even emergence of non-Fermi liquid states \cite%
{gegenwart2008quantum,QCP2020}.

Another way of enriching impurity dynamics is to make them mobile. Models of
mobile Fermi polarons were first considered in the context of He$^{4}$/He$%
^{3}$ mixtures, ions in the normal liquid of He$^{3}$, and diffusion of
muons in metals \cite{dMuon1998}. In comparison to the infinitely heavy
impurity models, a new feature of such systems comes from the finite recoil
energy of the impurity particle. This appears as a constraint on the
scattering processes of the bath fermions and raises the question of whether
the states with and without the impurity-bath coupling are orthogonal to
each other. For infinitely heavy impurities, we have the orthogonality
catastrophe, which means that the two states are orthogonal, while for heavy
but finite mass impurities, the answer was argued to depend on
dimensionality~\cite{rosch1999quantum}. In two- and three-dimensional
systems, the two states are expected to have a finite overlap, which in turn
implies a finite quasiparticle weight, whereas in one-dimensional systems,
the quasiparticle weight can be proven to vanish~\cite%
{castella1993,Dolgirev2020}.

It is interesting to note, however, that developing accurate theoretical
models for describing properties of mobile impurities interacting with a
Fermi bath remains a considerable theoretical challenge. Earlier studies of
mobile Fermi polarons have been motivated by two primary considerations. On
the one hand, they provided a concrete example of the emergence of friction
in a purely quantum-mechanical system~\cite{astrakharchik2004motion, cherny2012theory}. On the other hand, the issue of
quasiparticle weight was considered as a paradigmatic case study of the
concept of quasiparticles in strongly interacting Fermi systems.

Renewed interest in the study of Fermi polarons came with the progress of
experiments in the field of ultracold atoms. These systems make it possible
to realize Fermi polarons with different mass ratios of the impurity and
bath particles and tune impurity-bath interaction strength using magnetic
Feshbach resonances~\cite{schirotzek2009, schmidt2011excitation,zhang2012,kohstall2012, koschorreck_attractive_2012, cetina2016,scazza2017,yan2019,ness2020observation}. The tunability of microscopic interactions brings a new
feature of the interplay of few- and many-body aspects of the problem. In
particular, Feshbach resonance itself corresponds to the appearance of a
bound state in a two-body problem~\cite{chin2010}. An interesting question then is whether
one finds a transition between molecular and polaronic ground states in a
many-body system. In the former case, the impurity atom makes a bound state
with one of the bath fermions, accompanied by the vanishing quasiparticle
weight. In the polaronic case, the impurity interacts with many bath
particles and forms a state that has finite quasiparticle weight. In
three-dimensional systems, there is strong numerical \cite%
{prokofev2008a,prokofev2008b} and experimental \cite{schirotzek2009}
evidence for the polaron-to-molecule transition. Notably, one finds that in
the case of equal masses of impurity and bath particles, one can obtain a
good description of many-body polaronic states by including only a single
particle-hole excitation~\cite{chevy2006}. These so-called Chevy ansatz (CA)
wave functions work surprisingly well even at unitarity when the scattering
length diverges~\cite{combescot2007,combescot2008}. In two-dimensional (2D) systems, analysis based on CA
suggested that the ground state should always be of the polaronic type~\cite%
{zollner_polarons_2011} (for equal masses of the impurity and bath
fermions). However, analysis that extended CA to include two particle-hole
excitations supported the existence of the polaron-to-molecule transition~%
\cite{parish_polaron-molecule_2011, Parish2013, cui2020, Cheng2021}.

The most recent addition to the experimental platforms for exploring Fermi
polarons utilizes excitons and electrons (or holes) in transition metal
dichalcogenides (TMDs)~\cite{sidler2017}. In contrast to traditional Si and GaAs
semiconductors, TMDs have a smaller dielectric constant and heavier electron
mass, resulting in much stronger binding energy and smaller size of an
exciton. For a broad range of electron densities used in experiments, the
size of excitons is much smaller than a typical inter-electron distance.
Hence excitons can be treated as impurities when analyzing their interaction
with electrons. Furthermore, there is an effective Feshbach resonance
between electrons and excitons, which
manifests itself in the repulsive and attractive branches in the absorption
spectra. These branches are strongly reminiscent of the Fermi polaron
spectra measured in two-dimensional systems of ultracold fermions~\cite{sidler2017,efimkin2017,Christian2020,kuhlenkamp2022}.

Motivated by these developments, we set ourselves a goal of developing an
efficient theoretical formalism for describing Fermi polarons in 2D systems,
both in and out of equilibrium. The approach we choose here is based on the
non-Gaussian states detailed in Ref.~\cite{shi_variational_2018}. These
variational states do not rely on truncations in the number of particle-hole
excitations. As such, this approach provides an unbiased analysis of the
competition between polaronic and molecular states. It also guarantees that
in the limit of infinitely heavy impurities, our solution reduces to the
exact one based on the Functional Determinant Approach~\cite%
{levitov1996,knap2012,schmidt2018universal}. Additional motivation to employ the
non-Gaussian states is that they capture remarkably well the physics of 1D
Fermi polaron~\cite{mcguire_interacting_1965,mcguire_interacting_1966,mathy2012quantum, knap2014quantum, gamayun2016time,gamayun2018impact,gamayun2020zero,Dolgirev2020}. The main advantage of our method is the
possibility of analyzing non-equilibrium properties~\cite{knap2012,schmidt2012,parish2016quantum,schmidt2018universal,gamayun2018impact,liu2019variational,liu2020theory,adlong2020quasiparticle,burovski2021mobile} of polaronic systems,
including various spectral functions. As part of our analysis, we introduce
here new characteristics of Fermi-polaron systems, namely: molecular residue
and molecular spectral function. These quantities provide a complementary
perspective on the polaron-to-molecule transition. We discuss how they can
be measured in experiments with ultracold atoms.

This paper is organized as follows: In Sec.~\ref{sec: formalism}, we
introduce the general non-Gaussian approach, which includes the ground-state
optimization via the imaginary-time evolution, the study of the real-time
dynamics by projecting the Schr\"{o}dinger equation on the variational
manifold, and the linear-response analysis by linearizing the equations of
motion around the ground-state configuration. In Sec.~\ref{sec: static
properties}, a first-order polaron-to-molecule transition is identified,
where both the ground-state energy and single-particle residue are in
excellent agreement with those from CA and diagrammatic Monte Carlo (DMC)
calculations. Section~\ref{sec: dynamical properties} is dedicated to
far-from-equilibrium dynamics of the polaronic system. There, we compute
various spectral properties and introduce two types of RF spectroscopies to
quantify the polaron-to-molecule transition. The spectral functions exhibit
distinctive dynamical behaviors in the polaronic and molecular phases, such
as long-lived oscillations between the repulsive and attractive polarons and
fast relaxation to the molecular state from different initial states.
Finally, the main results are briefly summarized in Sec.~\ref{sec: summary}.

\section{\label{sec: formalism}Formalism}

This section introduces the non-Gaussian formalism to study the Fermi
polaron in two spatial dimensions. In the first subsection, we formulate the
model of a single impurity in the 2D Fermi gas and apply the Lee-Low-Pines
(LLP) transformation~\cite{LLP} that decouples the impurity from the
fermionic degrees of freedom. We then, in the second subsection, introduce
the non-Gaussian variational states, which allow us to investigate the
ground-state properties and the real-time dynamics. Up to this stage, our
framework closely follows that used in Ref.~\cite{Dolgirev2020} to study the
one-dimensional Fermi polaron. The two-dimensional problem is much more
challenging due to the large number of the involved degrees of freedom.
Consequently, numerical simulations are limited to small system sizes. To
overcome this issue, in the third subsection, we utilize the rotational
symmetry of the problem, which in turn allows us to efficiently model even
relatively large systems. The fourth subsection is dedicated to the linear
response theory within the formalism of the non-Gaussian states.

\subsection{Model}

A single mobile impurity immersed in a 2D Fermi gas is described via the
following microscopic Hamiltonian:
\begin{align}
H=H_{\mathrm{b}}+H_{\mathrm{imp}}+H_{\mathrm{int}},  \label{eqn:Ham_rs}
\end{align}
where
\begin{equation}
H_{\mathrm{b}}=-\frac{1}{2m_{\mathrm{b}}}\int d^{2}x \, \hat{\Psi}^{\dagger
}(\mathbf{x})\nabla ^{2}\hat{\Psi}(\mathbf{x})
\end{equation}%
represents the kinetic energy of the fermionic bath. The kinetic energy of
the impurity is given by:
\begin{equation}
H_{\mathrm{imp}}=-\frac{1}{2m_{\mathrm{imp}}}\int d^{2}x \, \hat{\Psi}_{%
\mathrm{imp}}^{\dagger }(\mathbf{x})\nabla ^{2}\hat{\Psi}_{\mathrm{imp}}(%
\mathbf{x}).
\end{equation}
The contact interaction term reads:
\begin{equation}
H_\mathrm{int}=g\int d^{2}x\, \hat{\Psi}^{\dagger }(\mathbf{x})\hat{\Psi}(%
\mathbf{x})\hat{\Psi}_{\mathrm{imp}}^{\dagger }(\mathbf{x})\hat{\Psi}_{%
\mathrm{imp}}(\mathbf{x}),  \label{eqn:Hint_rs}
\end{equation}%
where $\hat{\Psi}(\mathbf{x})$ and $\hat{\Psi}^{\dagger }(\mathbf{x})$ ($%
\hat{\Psi}_{\mathrm{imp}}(\mathbf{x})$ and $\hat{\Psi}_{\mathrm{imp}%
}^{\dagger }(\mathbf{x})$) denote the fermionic (impurity) creation and
annihilation operators, respectively; they obey the fermionic
anti-commutation relations. Here $m_{\mathrm{b}}$ and $m_{\mathrm{imp}}$ are
the fermion and impurity masses, respectively. In the 2D gas, the attractive
interaction strength $g$ is related to the 2D scattering length $a_{\mathrm{%
2D}}$ via the Lippmann-Schwinger equation:
\begin{equation}
\frac{1}{g}=-\frac{1}{L^2}\sum_{|\mathbf{k}|<k_{\Lambda }}\frac{1}{E_{B}+
k^2/(2m_{\mathrm{r}})},
\end{equation}
where $E_{B}=1/(2m_{\mathrm{r}}a_{\mathrm{2D}}^{2})$ is the binding energy
of the weakly bound diatomic molecule, $m_{\mathrm{r}}=m_{\mathrm{b}}m_{%
\mathrm{imp}}/(m_{\mathrm{b}}+m_{\mathrm{imp}})$ denotes the reduced mass, $L
$ is the linear system's size, and $k_{\Lambda }$ is the ultraviolet (UV)
momentum cutoff. In momentum space, the Hamiltonian reads:
\begin{align}
H = &\sum_{\mathbf{k}}(\varepsilon _{\mathrm{b},\mathbf{k}}c_{\mathbf{k}%
}^{\dagger }c_{\mathbf{k}} + \varepsilon _{\mathrm{imp},\mathbf{k}}f_{%
\mathbf{k}}^{\dagger }f_{\mathbf{k}})  \nonumber \\
&+\frac{g}{L^2}\sum_{\mathbf{k},\mathbf{p},\mathbf{q}}f_{\mathbf{k}%
}^{\dagger }f_{\mathbf{k}+\mathbf{q}}c_{\mathbf{p}}^{\dagger }c_{\mathbf{p-q}%
},
\end{align}
where $\varepsilon _{\sigma =\mathrm{b,imp},\mathbf{k}}=k^{2}/(2m_{\sigma })$
and $\mathbf{k}=2\pi (n_{x},n_{y})/L$ with integer $n_{x}$ and $n_{y}$.

We turn to discuss the Lee-Low-Pines (LLP) transformation $U_{\mathrm{LLP}%
}=e^{-i\mathbf{Q}_{b}\cdot \mathbf{X}_{\mathrm{imp}}}$, which allows one to
simplify the model by eliminating the impurity degrees of freedom. This
unitary transformation relies on the very fact of the total momentum
conservation $[\mathbf{Q}_{\mathrm{tot}},H]=0$, where $\mathbf{Q}_{\mathrm{%
tot}}= \mathbf{Q}_{\mathrm{imp }} + \mathbf{Q}_{\mathrm{b}}$. Here $\mathbf{Q%
}_{\mathrm{imp}}=\sum_{\mathbf{k}}\mathbf{k}f_{\mathbf{k}}^{\dagger }f_{%
\mathbf{k}}$ and $\mathbf{Q}_{\mathrm{b}}=\sum_{\mathbf{k}}\mathbf{k}c_{%
\mathbf{k}}^{\dagger }c_{\mathbf{k}}$ represent the momenta of the impurity
and the fermi bath, respectively. We also defined $\mathbf{X}_{\mathrm{imp}%
}=\int d^{2}x\,\mathbf{x}\,\hat{\Psi}_{\mathrm{imp}}^{\dagger }(\mathbf{x})%
\hat{\Psi}_{\mathrm{imp}}(\mathbf{x})$ to be the impurity position operator.
Physically, the LLP transformation simply encodes the fact that the impurity
momentum $\mathbf{Q}_{\mathrm{imp}}=U_{\mathrm{LLP}}^{\dagger }\mathbf{Q}_{%
\mathrm{tot}}U_{\mathrm{LLP}}$ can be reconstructed from the total momentum $%
\mathbf{Q}_{\mathrm{tot}}$ and the net momentum $\mathbf{Q}_{\mathrm{b}}$ of
the host fermions. Under the LLP transformation, the system is transformed
into the co-moving frame of the impurity. The modified Hamiltonian $H_{%
\mathrm{LLP}}=U_{\mathrm{LLP}}^{\dagger }HU_{\mathrm{LLP}}$ in the
single-impurity subspace $\sum_{\mathbf{k}}f_{\mathbf{k}}^{\dagger }f_{%
\mathbf{k}}=1$ then reads:%
\begin{align}
H_{\mathrm{LLP}} =& \sum_{\mathbf{k}}(\varepsilon _{\mathrm{b},\mathbf{k}}c_{%
\mathbf{k}}^{\dagger }c_{\mathbf{k}}+\varepsilon _{\mathrm{imp},\mathbf{k-Q}%
_{\mathrm{b}}}f_{\mathbf{k}}^{\dagger }f_{\mathbf{k}})  \nonumber \\
&+\frac{g}{L^2}\sum_{\mathbf{k}}f_{\mathbf{k}}^{\dagger }f_{\mathbf{k}}\sum_{%
\mathbf{p},\mathbf{q}}c_{\mathbf{p}}^{\dagger }c_{\mathbf{q}}.
\end{align}
We note that in the transformed frame, $f_{\mathbf{k}}^{\dagger }f_{\mathbf{k%
}}$ commutes with $H_{\mathrm{LLP}}$; in other words, $\mathbf{Q}_{\mathrm{%
imp}}=\mathbf{K}_{0}$ becomes an integral of motion in the co-moving frame
so that the LLP Hamiltonian can be written as:
\begin{equation}
H_{\mathrm{LLP}}=\sum_{\mathbf{k}}\varepsilon _{\mathrm{b},\mathbf{k}}c_{%
\mathbf{k}}^{\dagger }c_{\mathbf{k}}+\frac{g}{L^2}\sum_{\mathbf{k},\mathbf{p}%
}c_{\mathbf{k}}^{\dagger }c_{\mathbf{p}}+\varepsilon _{\mathrm{imp},\mathbf{K%
}_{0}\mathbf{-Q}_{\mathrm{b}}}.  \label{Decoupled Hamiltonian}
\end{equation}
We obtain that the impurity degrees of freedom are eliminated at the price
of introducing a non-local impurity-mediated interaction between the
fermions, encoded in the third term of Eq.~\eqref{Decoupled Hamiltonian}.

\subsection{\label{sec: variational method}Non-Gaussian variational approach}

\label{subsec:NGstates}

To study the polaron physics, both in and out of equilibrium, we employ the
non-Gaussian family of variational wave functions. Specifically, guided by
the LLP transformation, we write the many-body polaronic state in the
laboratory frame as:
\begin{align}
|\Psi _{\mathbf{K}_{0}}\rangle =U_{\mathrm{LLP}}f_{\mathbf{K}_{0}}^{\dagger
}|0\rangle_{\mathrm{imp}} \otimes |\Psi _{\mathrm{GS}}\rangle.
\label{eqn:NGS_wf}
\end{align}
Implicit in Eq.~\eqref{eqn:NGS_wf} is that the state $|\Psi _{\mathrm{GS}%
}\rangle$ represents the fermionic wave function in the co-moving frame. We
then choose $|\Psi _{\mathrm{GS}}\rangle$ to be Gaussian~\cite%
{shi_variational_2018,Dolgirev2020}:
\begin{align}
|\Psi _{\mathrm{GS}}\rangle =U_{\mathrm{GS}}|\mathrm{FS}\rangle =e^{-i\theta
}e^{ic^{\dagger }\xi c}|\mathrm{FS}\rangle,
\end{align}
where $|\mathrm{FS}\rangle$ describes the Fermi sea set by a Fermi momentum $%
k_F$. At this stage, our variational parameters are the global phase $\theta$
and Hermitian matrix $\xi$ written in the Dirac basis $c=\left( c_{\mathbf{k}%
_{1}},c_{\mathbf{k}_{2}},\ldots ,c_{\mathbf{k}_{N}}\right) ^{\mathrm{T}}$,
with $N$ being the total number of fermionic degrees of freedom. We note
that even though the wave function in the co-moving frame is factorizable
between the impurity and host fermions, it is highly entangled by $U_{%
\mathrm{LLP}}$ when expressed in the laboratory frame, cf. Eq.~%
\eqref{eqn:NGS_wf}.

Any variational state applied to many-body problems represents some
approximation. Given that often there are no small parameters or exact
solutions, it is crucial to test the validity of any such variational
approach. For the 1D Fermi polaron, it was demonstrated in Ref.~\cite%
{Dolgirev2020} that the non-Gaussian states of the form~\eqref{eqn:NGS_wf}
reproduce the exact Bethe ansatz results, both in and out of equilibrium.
The validity of the non-Gaussian wave functions in the 2D polaron problem is
the subject of the next sections.

To optimize for the best variational wave function that approximates the
ground state, we employ the imaginary-time dynamics. For now, instead of $%
\theta$ and $\xi$, it is more convenient to work with the covariance matrix
\begin{align}
\Gamma _{ij}\equiv \langle \Psi _{\mathrm{GS}}|c_{i}^{\dagger }c_{j}|\Psi _{%
\mathrm{GS}}\rangle =U^{\ast }\Gamma _{\mathrm{FS}}U^{\mathrm{T}},
\end{align}
where $\Gamma _{\mathrm{FS}}$ is the covariance matrix of the Fermi sea and $%
U = e^{i\xi}$. Then the projection of the imaginary-time evolution onto the
tangential space of the variational manifold gives rise to~\cite%
{shi_variational_2018}:
\begin{equation}
\partial _{\tau }\Gamma =-\mathcal{H}_{\mathrm{MF}}^{\mathrm{T}}\Gamma
-\Gamma \mathcal{H}_{\mathrm{MF}}^{\mathrm{T}}+2\Gamma \mathcal{H}_{\mathrm{%
MF}}^{\mathrm{T}}\Gamma.
\end{equation}
Here we employed Wick's theorem to derive the mean-field Hamiltonian%
\begin{align}
(\mathcal{H}_{\mathrm{MF}})_{\mathbf{kp}} =& \left(\frac{k^{2}}{2m_{\mathrm{r%
}}}\mathbf{-}\frac{\mathbf{K}_{0}\cdot \mathbf{k}}{m_{\mathrm{imp}}}%
\right)\delta _{\mathbf{kp}}+\frac{g}{L^2} \\
&+\frac{1}{m_{\mathrm{imp}}}\left(\left\langle \mathbf{Q}_{\mathrm{b}%
}\right\rangle_{\mathrm{GS}} \cdot \mathbf{k}\, \delta _{\mathbf{kp}}-%
\mathbf{k\cdot p}\left\langle c_{\mathbf{p}}^{\dagger }c_{\mathbf{k}%
}\right\rangle_{\mathrm{GS}} \right).  \nonumber
\end{align}
For the imaginary-time evolution, the global phase $\theta $ can be chosen
arbitrarily, and the variational energy%
\begin{align}
E_{\mathbf{K}_{0}} =&\mathrm{Tr}(\mathcal{H}_{\mathrm{MF}}\Gamma ^{\mathrm{T}%
})+\frac{1}{2m_{\mathrm{imp}}}\Big(K_{0}^{2}+\left\langle \mathbf{Q}_{%
\mathrm{b}}\right\rangle^{2}_{\mathrm{GS}}  \nonumber \\
&-\sum_{\mathbf{kp}}\mathbf{k\cdot p} \langle c_{\mathbf{k}}^{\dagger}c_{%
\mathbf{p}}\rangle _{\mathrm{GS}} \langle c_{\mathbf{p}}^{\dagger }c_{%
\mathbf{k}}\rangle_{\mathrm{GS}} \Big)  \label{Energy function}
\end{align}%
decreases monotonically and reaches its ground-state value in the limit $%
\tau \rightarrow \infty $.

The real-time equations of motion are derived from Dirac's variational
principle, with the result~\cite{Dolgirev2020}:
\begin{eqnarray}
\partial _{t}U &=&-i\mathcal{H}_{\mathrm{MF}}U, \\
\partial _{t}\theta  &=&E_{\mathbf{K}_{0}}-\mathrm{Tr}\left( \mathcal{H}_{%
\mathrm{MF}}\Gamma ^{\mathrm{T}}\right) .
\end{eqnarray}%
From this, one can get an equation solely on the covariance matrix:
\begin{equation}
i\partial _{t}\Gamma =\Gamma \mathcal{H}_{\mathrm{MF}}^{\mathrm{T}}-\mathcal{%
H}_{\mathrm{MF}}^{\mathrm{T}}\Gamma .
\label{eq:real-time evolution of the Gamma matrix}
\end{equation}%
This result could alternatively be derived from projecting the Schr\"{o}%
dinger equation onto the tangential space of the variational manifold~\cite%
{shi_variational_2018}. 

As a remark, we note that during either the imaginary-time or real-time evolution, the total number of fermions is conserved $d_{\tau ,t}N_{f}{=0}
= d_{\tau ,t}\mathrm{Tr}(\Gamma )=0$. This follows from the fact that provided the initial state is pure, as encoded in $\Gamma^2 = \Gamma$, it will remain pure upon the evolution.

\subsection{Rotational symmetry}

Let us consider the situation with zero total momentum $\mathbf{K}_{0}=%
\mathbf{0}$, where the system is rotationally invariant. When performing the
LLP transformation for this case, we work with continuous rather than
discretized variables, as in Eqs.~\eqref{eqn:Ham_rs}-\eqref{eqn:Hint_rs}.
The rotational symmetry implies that the covariance matrix $\Gamma _{\mathbf{%
pp}^{\prime }}$ (or any other observable) depends only on $p$, $p^{\prime }$%
, and the relative angle $\vartheta -\vartheta ^{\prime }$, which allows us
to write:
\begin{equation}
\Gamma _{\mathbf{pp}^{\prime }}=\frac{1}{2\pi \delta _{p}\sqrt{pp^{\prime }}}%
\sum_{n}\Gamma _{pp^{\prime }}^{n}e^{-in(\vartheta -\vartheta ^{\prime })}.
\end{equation}%
In this expression, the radial momenta $p$ and $p^{\prime }$ in each of the
matrices $\Gamma _{pp^{\prime }}^{n}$ have been discretized with spacing $%
\delta _{p}$. Here, $\Gamma _{pp^{\prime }}^{n}$ is understood as the following
covariance matrix:
\begin{equation}
\Gamma _{pp^{\prime }}^{n}=\langle \Psi _{\mathrm{GS}}|c_{p}^{n\dagger
}c_{p^{\prime }}^{n}|\Psi _{\mathrm{GS}}\rangle ,
\end{equation}%
where $c_{p}^{n}\propto \int d\vartheta \,c_{\mathbf{p}}e^{-in\vartheta }$
satisfing $[c_{p}^{n},c_{p^{\prime }}^{n\dagger }]=\delta _{pp^{\prime }}$
is the annihilation operator in the angular momentum basis. The
imaginary-time equations of motion now read:
\begin{equation}
\partial _{\tau }\Gamma ^{n}=-[\mathcal{H}^{n}]^{\mathrm{T}}\Gamma
^{n}-\Gamma ^{n}[\mathcal{H}^{n}]^{\mathrm{T}}+2\Gamma ^{n}[\mathcal{H}%
^{n}]^{\mathrm{T}}\Gamma ^{n},
\end{equation}%
where the mean-field Hamiltonian in the angular momentum channel $n$ is
given by:
\begin{align}
\mathcal{H}_{pp^{\prime }}^{n}=& \frac{p^{2}}{2m_{\mathrm{r}}}\delta
_{pp^{\prime }}+\frac{g\delta _{p}\sqrt{pp^{\prime }}}{2\pi }\delta _{n0}
\nonumber \\
& -\frac{pp^{\prime }}{2m_{\mathrm{imp}}}\left( \Gamma _{p^{\prime
}p}^{n+1}+\Gamma _{p^{\prime }p}^{n-1}\right) .  \label{eqn H_n}
\end{align}%
As encoded in the second term in Eq.~\eqref{eqn H_n}, the impurity induces a
potential in the zero angular momentum channel only -- this is because we
consider contact coupling. We note that eventually the distribution of
fermions for $n\neq 0$ becomes affected via the inter-channel scattering
described by the third term in Eq.~\eqref{eqn H_n}. The real-time evolution
for the unitary $U^{n}=\exp \left( i\xi ^{n}\right) $ in the channel $n$ and
the global phase $\theta $ read:%
\begin{eqnarray}
\partial _{t}U^{n} &=&-i\mathcal{H}^{n}U^{n}, \\
\partial _{t}\theta  &=&E_{0}-\overset{\infty }{\sum_{n=-\infty }}\mathrm{Tr}%
\left( \mathcal{H}^{n}\Gamma ^{n\mathrm{T}}\right) ,  \label{EOMn}
\end{eqnarray}%
where the energy functional $E_{0}$ for $\mathbf{K}_{0}=\mathbf{0}$ depends
on each $\Gamma ^{n}$ and is expressed as:
\begin{align}
E_{0}=& \sum_{p,n}\frac{p^{2}}{2m_{\mathrm{r}}}\Gamma _{pp}^{n}+\frac{%
g\delta _{p}}{2\pi }\sum_{p,p^{\prime }}\,\sqrt{pp^{\prime }}\,\Gamma
_{pp^{\prime }}^{0}  \nonumber \\
& -\frac{1}{2m_{\mathrm{imp}}}\sum_{p,p^{\prime },n}pp^{\prime }\,\Gamma
_{pp^{\prime }}^{n}\Gamma _{p^{\prime }p}^{n+1}.
\end{align}%
The main result of this subsection is that the initial two-dimensional
problem reduces to simulating coupled one-dimensional ones, which
dramatically facilitates numerical analyses of even relatively large
systems. In practice, we introduce a cutoff $n_{\Lambda }$ in angular
momentum space such that the covariance matrix for $|n|>n_{\Lambda }$ is
replaced by the expectation value for the filled Fermi sea $\Gamma
_{pp^{\prime }}^{n}=\delta _{pp^{\prime }}\theta \left( k_{F}-p\right) $.
The value $n_{\Lambda }$ is determined by the numerical convergence of the
results. We remark that for the Gaussian state considered in this
subsection, different angular momentum channels are decoupled, and the
particle number of each channel is individually conserved, i.e., $d_{\tau
,t}N_{f}^{n}=d_{\tau ,t}\mathrm{Tr}(\Gamma ^{n})=0$.

\subsection{Linear response formalism}

\label{subsec_CM_analysis}

One of the goals of this work is to provide a framework capable of computing
observables relevant for both solid-state and ultracold atoms experiments.
In this subsection, we focus on linear-response probes, which in turn
require careful analysis of collective modes representing small-amplitude
fluctuations on top of a (momentum-dependent) ground state. We remark that
the fluctuation analysis within Gaussian states for a bosonic system has
been proven to be equivalent to the generalized random phase approximation
and successfully applied to reproduce the Goldstone zero-mode, naturally
without imposing the Hugenholtz-Pines condition~\cite{demler1996class,
Guaita2019,shi2019trapped}. For the 1D Fermi polaron, collective modes
turned out to be crucial for understanding even far-from-equilibrium
properties~\cite{Dolgirev2020}.

In the LLP frame, the particle-hole excitation spectrum can be
analyzed via linearizing Eq.~\eqref{eq:real-time evolution of the Gamma
matrix} around the ground-state configuration, characterized by $U_{g}=%
\mathrm{exp}(i\Xi _{g})$ and $\Gamma _{g}=U_{g}^{\ast }\Gamma _{0}U_{g}^{%
\mathrm{T}}$. We note that the unitary $U_{g}$ diagonalizes the mean-field
Hamiltonian $U_{g}^{\dagger }\mathcal{H}_{\mathrm{MF}}U_{g}=d_{g}$.
Small-fluctuations $\mathrm{\delta }\Xi$ are encoded in the fermionic wave
function as:
\begin{equation}
e^{i\hat{c}^{\dagger }\Xi _{g}\,\hat{c}}e^{i\hat{c}^{\dagger }\mathrm{\delta
}\Xi \hat{c}}|\mathrm{FS}\rangle,
\end{equation}
where the particle-hole generator $\mathrm{\delta }\Xi $ is an $N\times N$
Hermitian matrix ($N$ is the total number of single-particle modes in the
fermionic system). The corresponding unitary matrix becomes $U=U_{g}e^{i%
\mathrm{\delta }\Xi }$. The gauge redundancy in $\mathrm{\delta }\Xi $ can
be eliminated by requiring the non-vanishing fluctuation $\mathrm{\delta }%
\Gamma \equiv \Gamma -\Gamma _{g}\sim -iU_{g}^{\ast }[\mathrm{\delta }\Xi
^{\ast },\Gamma _{0}]U_{g}^{\mathrm{T}}$ of the covariance matrix. Since the
covariance matrix of the state $|\mathrm{FS}\rangle $ composed of $N_f$
fermions is $\Gamma_\mathrm{FS}=\left(
\begin{array}{cc}
\mathrm{I}_{N_f\times N_f} & 0 \\
0 & 0%
\end{array}%
\right) $, the condition $\mathrm{\delta }\Gamma \neq 0$ imposes the
off-diagonal form $\mathrm{\delta }\Xi =\left(
\begin{array}{cc}
0 & \mathrm{\delta }\xi \\
\mathrm{\delta }\xi ^{\dagger } & 0%
\end{array}%
\right) $ with an $N_f\times (N-N_f)$ matrix $\mathrm{\delta }\xi$. In terms
of $\mathrm{\delta }\xi$, the fluctuation of the covariance matrix reads%
\begin{equation}
\mathrm{\delta }\Gamma =U_{g}^{\ast }\left(
\begin{array}{cc}
0 & i\mathrm{\delta }\xi ^{\ast } \\
-i\mathrm{\delta }\xi ^{\mathrm{T}} & 0%
\end{array}%
\right) U_{g}^{\mathrm{T}}.
\end{equation}%
Linearization of Eq.~(\ref{eq:real-time evolution of the Gamma matrix})
results in%
\begin{equation}
i\partial _{t}\mathrm{\delta }\Xi =[d_{g},\mathrm{\delta }\Xi
]-iU_{g}^{\dagger }\mathrm{\delta }\mathcal{H}U_{g},
\label{eq: linearized EOM of xi}
\end{equation}%
where the fluctuation matrix $%
\mathrm{\delta }\mathcal{H}$ describing particle-hole interactions is given by:
\begin{equation}
\delta \mathcal{H}_{\mathbf{kp}}=\delta_{\mathbf{k} \mathbf{p}} \sum_{%
\mathbf{q}} \frac{\mathbf{k} \cdot \mathbf{q}}{M} \mathrm{\delta }\Gamma_{%
\mathbf{qq}}-\frac{\mathbf{k} \cdot \mathbf{p}}{M} \delta \Gamma_{\mathbf{pk}%
}.
\end{equation}
Provided $\mathbf{K}_{0}=\mathbf{0}$%
, following the preceding subsection, we write $\delta \mathcal{H}$ as:
\begin{align}
\mathrm{\delta }\mathcal{H}_{pn,p^{\prime }n^{\prime }} =&\frac{p\delta
_{pp^{\prime }}}{2m_{\mathrm{imp}}}\sum_{q, m,\sigma =\pm 1}q\delta
_{n^{\prime }n+\sigma }\mathrm{\delta }\Gamma _{qm+\sigma ,qm}  \nonumber \\
& -\frac{pp^{\prime }}{2m_{\mathrm{imp}}}\sum_{\sigma =\pm 1}\mathrm{\delta }%
\Gamma _{p^{\prime }n^{\prime }+\sigma ,pn+\sigma }.
\end{align}%
Equation~\eqref{eq: linearized EOM of xi} gives rise to a compact equation
of motion $i\partial _{t}v_{\mathrm{ph}}=\mathcal{M}v_{\mathrm{ph}}$, where $%
v_{\mathrm{ph}}=\left( \mathrm{\delta }\xi ,\mathrm{\delta }\xi ^{\ast
}\right) ^{T}$. The spectrum of collective modes is given by the eigenvalues
of $\mathcal{M}$; linear-response observables also require the knowledge of
the eigenvectors of $\mathcal{M}$. We finally remark that for $\mathbf{K}%
_{0}=\mathbf{0}$, one can write $\xi _{pn,pn^{\prime }}=\delta _{nn^{\prime
}}\delta \xi _{pp^{\prime }}^{|n|}$, which further facilitates numerical
evaluations. An example of analysis of collective modes is discussed in
Appendix~\ref{appendix:collective modes}.

\section{\label{sec: static properties}Ground-state properties}

\begin{figure}[t!]
\includegraphics[width=1\linewidth]{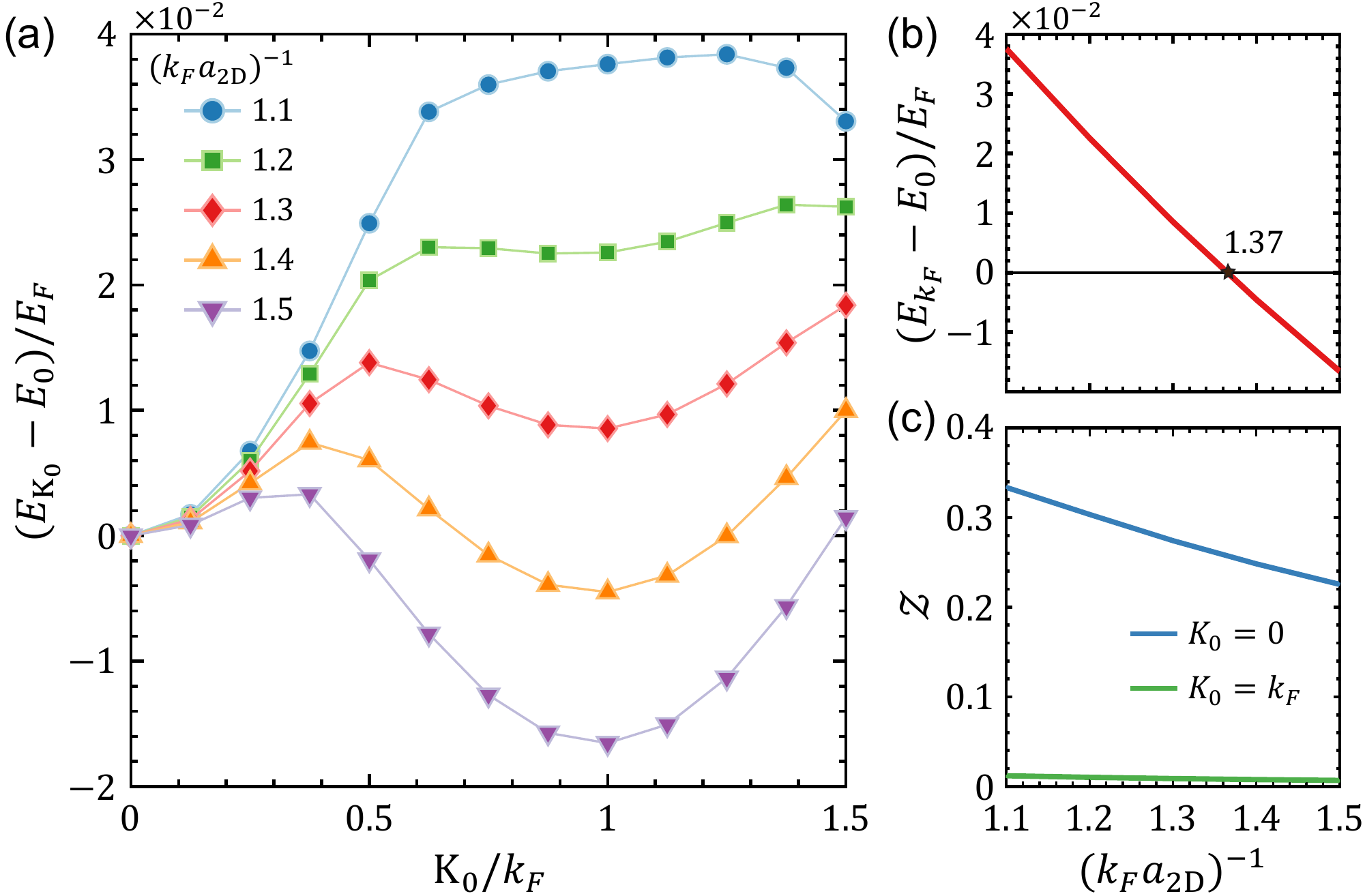}
\caption{Polaron-to-molecule phase transition within the non-Gaussian
variational states. (a) Polaron energy-momentum relation for various values
of $(k_F a_{\mathrm{2D}})^{-1}$: as this parameter is increased, we observe
that the minimum in the dispersion shifts from being located at $\mathrm{K}%
_0 = 0$ to $\mathrm{K}_0 = k_F$, indicating a first-order
polaron-to-molecule transition. Here $E_F = k_F^2/2m_{\mathrm{b}}$ is the
Fermi energy. (b) This transition is predicted to occur at around $(k_F a_{%
\mathrm{2D}})^{-1} = 1.37$. (c) The quasiparticle residue $\mathcal{Z}$ for $%
\mathrm{K}_0 = 0$ remains finite as the transition point is crossed; the
residue for $\mathrm{K}_0 = k_F$ is close to zero. Parameters used: $m_{%
\mathrm{imp}}=5m_{\mathrm{b}},k_{\Lambda }=5k_{F}$, and $\protect\delta %
_{p}=2\protect\pi /L=\left. k_{F}\right/8$. }
\label{fig 1}
\end{figure}

In this section, we primarily investigate the polaron-to-molecule phase
transition. We begin by exploring the full polaron energy-momentum relation,
being interested in arbitrary total momentum $\mathbf{K}_0$. As such, the
system is, in general, not rotationally symmetric, and numerical simulations
are computationally expensive. To facilitate the computations, we consider,
for now, rather heavy impurities, such as $m_{\mathrm{imp}}=5m_{\mathrm{b}}$%
, allowing us to choose a rather small UV cutoff $k_{\Lambda}$ because of
relatively small binding energy $E_B$ -- this energy decreases with
increasing the ratio $m_{\mathrm{imp}}/m_{\mathrm{b}}$~\cite%
{parish_polaron-molecule_2011}. If one is interested solely in the case with
$\mathrm{K}_0 = 0$, the polaronic properties can be efficiently studied for
arbitrary mass ratios using rotational symmetry, as we discuss below.

Figure~\ref{fig 1}(a) shows the polaron energy-momentum relation for various
interaction strengths, as encoded in the dimensionless parameter $\left(
k_{F}a _{\mathrm{2D}}\right) ^{-1}$. We note that this dispersion $E_{%
\mathbf{K}_{0}}$ depends on $\mathrm{K}_0$ only (it does not depend on the
direction of ${\mathbf{K}_{0}}$). Notably, for sufficiently strong
interactions, the energy of the state at $\mathrm{K}_0 = k_F$ becomes
smaller than that at $\mathrm{K}_0 = 0$, indicating a change in the nature
of the ground state -- this change occurs at around $\left( k_{F}a _{\mathrm{%
2D}}\right) ^{-1} = 1.37$, as shown in Fig.~\ref{fig 1}(b). To better
understand this transition, we now consider the quasiparticle residue
defined as:
\begin{align}
\mathcal{Z} = \left\vert \langle \mathrm{FS}|f_{\mathbf{K}_{0}}|\Psi _{%
\mathbf{K}_{0}}\rangle \right\vert ^{2}.
\end{align}
This expression can be understood as the overlap between the non-interacting
many-body state $f_{\mathbf{K}_{0}}^{\dagger }|\mathrm{FS}\rangle$ and the
true ground state $|\Psi _{\mathbf{K}_{0}}\rangle$ with the impurity-bath
interaction being switched on. Within the non-Gaussian states, the polaron residue is given by~\cite{Dolgirev2020}:
\begin{align}
\mathcal{Z}=\left\vert \langle \mathrm{FS}|\Psi _{\mathrm{GS}}\rangle
\right\vert^{2} =\det\left(\mathrm{I}_{N}+2 \Gamma_\mathrm{FS}
\Gamma-\Gamma-\Gamma_\mathrm{FS}\right).
\end{align}
Figure~\ref{fig 1}(c) shows the quasiparticle residues at $\mathrm{K}_0 = 0$
and $\mathrm{K}_0 = k_F$ across the transition: while the former smoothly
decreases with $\left( k_{F}a _{\mathrm{2D}}\right) ^{-1}$ and remains
finite at the transition point, the latter is nearly zero. We remark that these results agree with the studies in  Refs.~\cite{cui2020, Cheng2021}.

\begin{figure}[t!]
\includegraphics[width=1\linewidth]{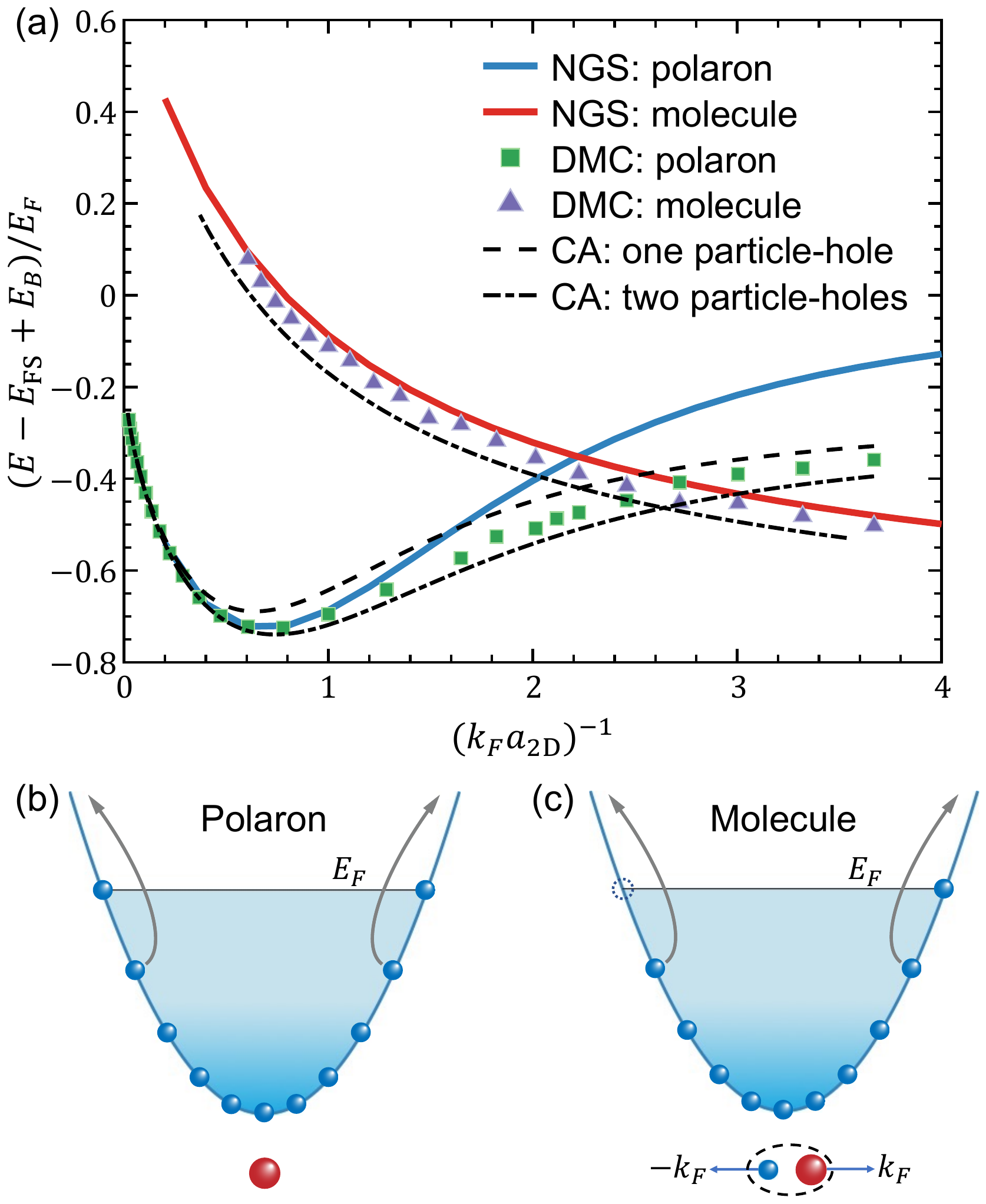}
\caption{ Comparison to the existing methods. (a) Energies of the polaronic
and molecular states as a function of $\left( k_{F}a_{\mathrm{2D}}\right)
^{-1}$ for the case of equal masses $m_{\mathrm{imp}}=m_{\mathrm{b}}$. The
polaron-to-molecule transition point within the NGS approach is around $%
\left( k_{F}a_{\mathrm{2D}}\right) ^{-1}=2.2$. For comparison, we also show
the known results from DMC~\protect\cite{DMC2D} and CA with one and two
particle-hole excitations~\protect\cite{Parish2013}. Here $E_{\mathrm{FS}}$
is the energy of the filled Fermi sea. Parameters used: $k_{\Lambda }=20k_{F}
$, $\protect\delta_{p}= k_{F}/ 40$, and $n_{\Lambda }=8$. (b) Schematic of
the polaronic state $\mathbf{K}_{0} = (0,0)$: the impurity (red ball) has on average zero momentum
and is coupled to particle-hole excitations (shown with arrows) of the Fermi
sea (fermions are shown as blue balls). (c) Schematic of the molecular
state $\mathbf{K}_{0} = (k_F,0)$: the mobile impurity with momentum around $(k_F,0)$ binds to the fermionic atom that has momentum $(-k_F,0)$; the resulting molecule on average has zero
momentum and weakly interacts with the rest of the Fermi sea.}
\label{fig 2}
\end{figure}

These findings suggest the following physical picture. For weak and moderate
interactions, the ground state is polaronic, it corresponds to the solution
with $\mathrm{K}_0 = 0$, and it has finite quasiparticle weight [Fig.~\ref%
{fig 2}(b)]. For stronger interactions, the system exhibits a first-order
phase transition into a molecular state, associated with the solution with $%
\mathrm{K}_0 = k_F$ and vanishing quasiparticle residue $\mathcal{Z} = 0$.
In this regime, we find that for $\mathbf{K}_{0}=(k_{F},0)$, the fermion
occupation at $(-k_{F},0)$ is essentially zero, indicating that this fermion
has been removed from the Fermi sea to form a bound state with the impurity
particle, so that the resulting molecule approximately has zero net momentum
[Fig.~\ref{fig 2}(c)]. We finally remark that if one would have limited the
analysis only to the $\mathrm{K}_0 =0$ sector, instead of an abrupt
transition, one would find a smooth crossover with gradual suppression of
the quasiparticle weight.

When investigating the polaron-to-molecule transition for lighter
impurities, such as $m_{\mathrm{imp}}=m_{\mathrm{b}}$, the binding energy $%
E_{B}$ becomes large, requiring a larger UV cutoff $k_\Lambda$ and making
computations too expensive. We now argue that rotational symmetry can be
naturally used to overcome this difficulty. The analysis of the polaronic
state is immediately simplified because this state corresponds to $\mathbf{K}%
_0 = (0,0)$, where the system is already rotationally invariant. For the
molecular state, we have $\mathbf{K}_{0}=(k_{F},0)$ and, thus, rotational
symmetry is broken. To restore this symmetry in our variational ansatz, we
employ the following method: instead of working with $N_f$ fermions in the
sector $\mathbf{K}_{0}=(k_{F},0)$, we add one more extra fermion and work in
the sector $\mathbf{K}_{0}=(0,0)$. In other words, to describe the molecular
state, from now on we will use the following variational wave function:
\begin{align}
|\Psi _{\mathrm{NGS}}^{N_f+1}\rangle =U_{\mathrm{LLP}}f_{\mathbf{K}_{0} =
0}^{\dagger }|\Psi _{\mathrm{GS}}^{N_f+1}\rangle,
\end{align}
where $|\Psi _{\mathrm{GS}}^{N_f+1}\rangle$ is chosen to be a Gaussian state
for $N_f+1$ fermions. This insight comes from our previous observation that
the impurity tends to bind one of the Fermi surface fermions -- see Fig.~\ref%
{fig 2}(c). Thus, the newly introduced fermion fills the hole in the
disturbed Fermi sea and makes the total momentum of the enlarged system
zero. One can alternatively view the simplified molecular state in the
spirit of Yosida's ansatz~\cite{Yosida1966}, where one has a Fermi sea of $%
N_f$ fermions, and the impurity and extra bath fermion form a bound state
with zero net momentum. In this molecule, both the impurity and extra
fermion have to be outside the Fermi sea because their momenta should be
opposite, but the extra fermion is excluded from the Fermi sea by the Pauli
principle. We emphasize that our variational state goes beyond this simple
ansatz because we take into account particle-hole excitations of the Fermi
sea arising from the nonzero impurity-bath coupling.

\begin{figure}[t]
\includegraphics[width=1\linewidth]{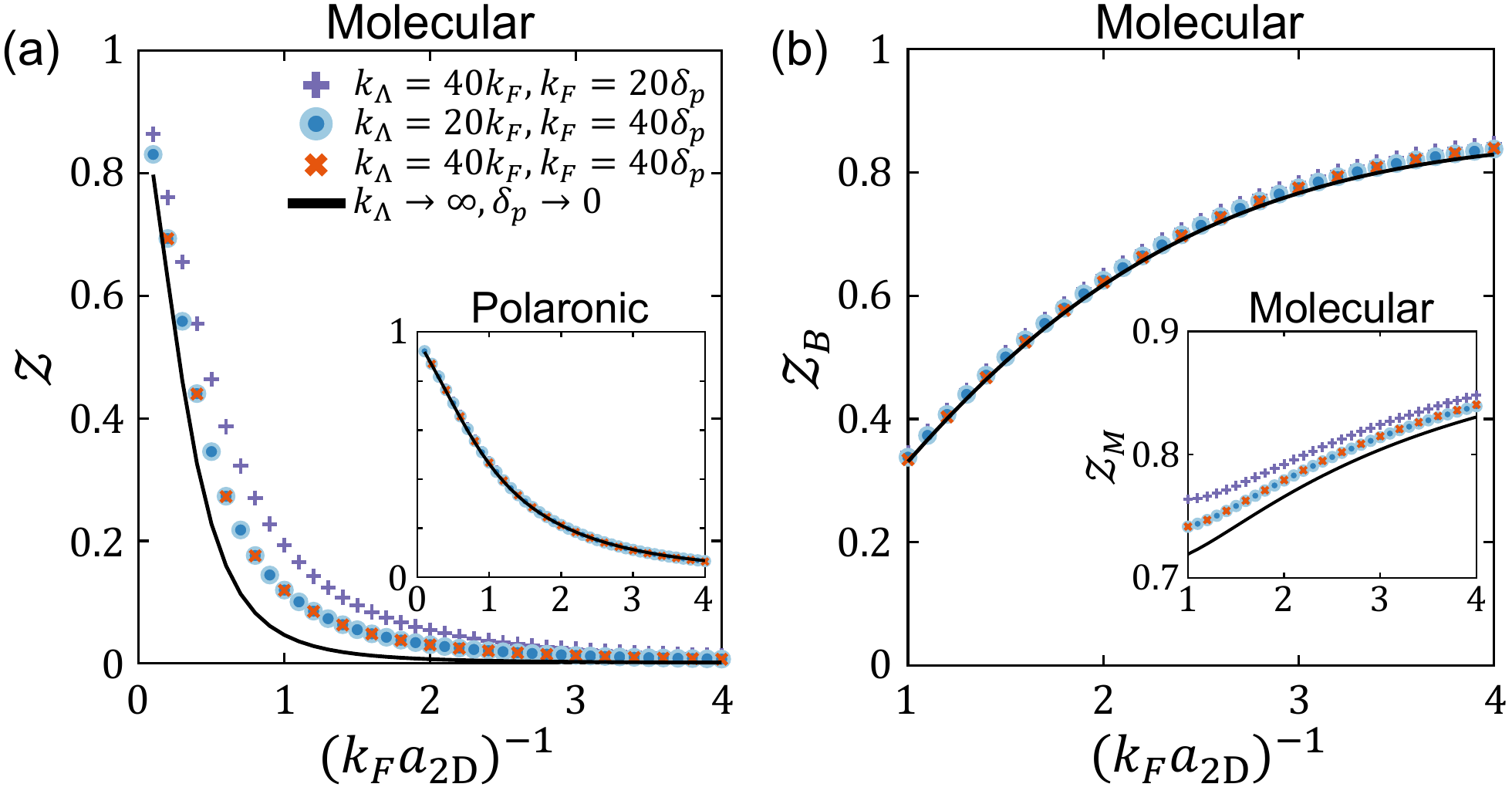}
\caption{Various residues and their convergence with varying the UV cutoff $%
k_\Lambda$ and the linear system's size encoded in $\protect\delta_p$. Solid
lines denote the extrapolation to the continuum limit $k_{\Lambda
}\rightarrow \infty $ and $\protect\delta_{p}\rightarrow 0$. Parameters are
the same as in Fig.~\protect\ref{fig 2}.}
\label{fig 3}
\end{figure}

Figure~\ref{fig 2}(a) shows the energies of polaronic and molecular states
as functions of $\left( k_{F}a_{\mathrm{2D}}\right) ^{-1}$ for $m_{\mathrm{%
imp}}=m_{\mathrm{b}}$. We find that our non-Gaussian variational approach
quantitatively agrees with the known results from the CA~\cite{Parish2013}
and DMC calculations~\cite{DMC2D}. Our method is particularly accurate at
capturing the molecular branch, confirming the validity of the simplified
molecular ansatz $|\Psi _{\mathrm{NGS}}^{N_{f}+1}\rangle $. The
polaron-to-molecule transition is predicted to occur around $\left( k_{F}a_{%
\mathrm{2D}}\right) ^{-1}=2.2$. Across the transition point, the polaronic
residue $\mathcal{Z}$ remains finite [inset of Fig.~\ref{fig 3}(a)], but the
molecular one is essentially zero [Fig.~\ref{fig 3}(a)]. Accurate analysis
of the convergence of our results with the UV cutoff $k_{\Lambda }$ and the
parameter $\delta _{p}$ that encodes the linear system's size indicates that
indeed, in the limit $k_{\Lambda }\rightarrow \infty $ and $\delta
_{p}\rightarrow 0$, the molecular residue approaches zero for $\left(
k_{F}a_{\mathrm{2D}}\right) ^{-1}\gtrsim 2.2$ -- see the solid line in Fig.~%
\ref{fig 3}(a). For $\left( k_{F}a_{\mathrm{2D}}\right) ^{-1}<2.2$, the
ground state is polaronic, and, as such, the molecular state corresponds to some excited
state. Since the size of this molecule becomes more extensive as $\left( k_{F}a_{%
\mathrm{2D}}\right) ^{-1}$ is decreased, accurate computation of the residue $\mathcal{Z}$ for
small $\left( k_{F}a_{\mathrm{2D}}\right) ^{-1}$ requires a smaller infrared
cutoff $\delta _{p}$.

Finally, we finish this section by introducing two more ``molecular residues" that help characterize the molecular state better. Since the
quasiparticle residue is close to being one in the polaronic phase and
vanishes in the molecular phase, the new molecular residues should display
the opposite behavior. Motivated by this, we introduce the first one as $%
\mathcal{Z}_{M}=|\langle \Psi_{M}|\Psi _{\mathrm{NGS}}^{N_f+1}\rangle |^{2}$%
, where $|\Psi_{M}\rangle =\sum_{\left\vert \mathbf{k}\right\vert
>k_{F}}\varphi _{\mathbf{k}}^{M}c_{\mathbf{k}}^{\dagger }f_{-\mathbf{k}%
}^{\dagger }\left\vert \mathrm{FS}\right\rangle$ encodes the Yosida ansatz.
Optimization of the variational parameters gives $\varphi _{\mathbf{k}%
}^{M}\propto -1/(k^{2}+a_\mathrm{2D}^{-2}-k_{F}^{2})$. We define the second residue as $\mathcal{Z}%
_{B}=|\langle \Psi_{B}|\Psi _{\mathrm{NGS}}^{N_f+1}\rangle |^{2}$, where $%
|\Psi _{B}\rangle =\sum_{\left\vert \mathbf{k}\right\vert >k_{F}}\varphi _{%
\mathbf{k}}^{B}c_{\mathbf{k}}^{\dagger }f_{-\mathbf{k}}^{\dagger }\left\vert
\mathrm{FS}\right\rangle $ with the parameters $\varphi _{\mathbf{k}}^{B}$
given by:
\begin{equation}
\varphi _{\mathbf{k}}^{B}=\frac{1}{ L\sqrt{\mathcal{N}}}\frac{1}{- 1/(2 m_{%
\mathrm{r}} a_X^2) -\varepsilon _{\mathrm{imp},\mathbf{k}}-\varepsilon _{%
\mathrm{b},\mathbf{k}}},  \label{B-type molecular wavefunction}
\end{equation}
Here, $\mathcal{N}=m_{\mathrm{r}}^{2}a_X^{2}/[\pi (1+k_{F}^{2}a_X^{2})]$ is
a normalization constant. The residue $\mathcal{Z}_{B}$ is useful because it can be measured with
ultracold atom experiments, as we elaborate in the next section. We also
postpone the discussion of the newly introduced length $a_{X}$ to the next
section but will assume here that it is much smaller than the Fermi
wavelength. We only emphasize here that $a_{X}$ is different from
the scattering length $a_{\mathrm{2D}}$.

We observe that the states $|\Psi_{B,M}\rangle$ can be written as $%
|\Psi_{B,M}\rangle =U_{\mathrm{LLP}}f_\mathbf{0}^{\dagger}|0\rangle_{\mathrm{%
imp}} \otimes|\bar{\Psi}_{B,M}\rangle$, with the Gaussian states $|\bar{\Psi}_{B,M}\rangle$ given by:
\begin{equation}
|\bar{\Psi}_{B,M}\rangle= \sum_{\left\vert \mathbf{k}\right\vert
>k_{F}}\varphi^{B,M} _{\mathbf{k}}c_{\mathbf{k}}^{\dagger }\left\vert
\mathrm{FS}\right\rangle.  \label{eq: psibar B}
\end{equation}
The molecular residues $\mathcal{Z}_{B,M}$ are then computed through the corresponding covariance matrices  $\Gamma_{B,M}$ as:
\begin{align}
\mathcal{Z}_{B,M}&=\left\vert \langle \bar{\Psi}_{B,M}|\Psi _{\mathrm{GS}
}\rangle \right\vert^{2}  \nonumber \\
&=\det\left(\,\mathrm{I}_{N}+2
\Gamma_{B,M}\Gamma-\Gamma-\Gamma_{B,M}\right).
\end{align}

Figure~\ref{fig 3}(b) shows the dependence of the two residues on $(k_F a_{%
\mathrm{2D}})^{-1}$. Both of them approach one in the molecular phase.
Physically, deep inside the molecular phase, the impurity and one bath
fermion form a tight bare bound state, that in turn creates a scattering
potential to the rest of the bath particles. Close to the phase transition,
the bath fermions start to strongly affect the structure of the bound state,
resulting in, for instance, a small overlap $\vert \langle \Psi_{B}|\Psi _{%
\mathrm{NGS}}^{N_f+1}\rangle \vert $. For the residue $\mathcal{Z}_M$, we
find that even though the parameters $\varphi _{\mathbf{k}}^{M}$ are being
optimized, $\mathcal{Z}_{M}\sim 0.8$ is still smaller than one in the
molecular phase. We attribute this deviation to the fact that the Yosida
ansatz, in contrast to the non-Gaussian wave function, does not take into
account the backreaction from the Fermi sea on the formation of the
molecular bound state.

\section{\label{sec: dynamical properties}Dynamical properties}

\begin{figure}[t!]
\includegraphics[width=1\linewidth]{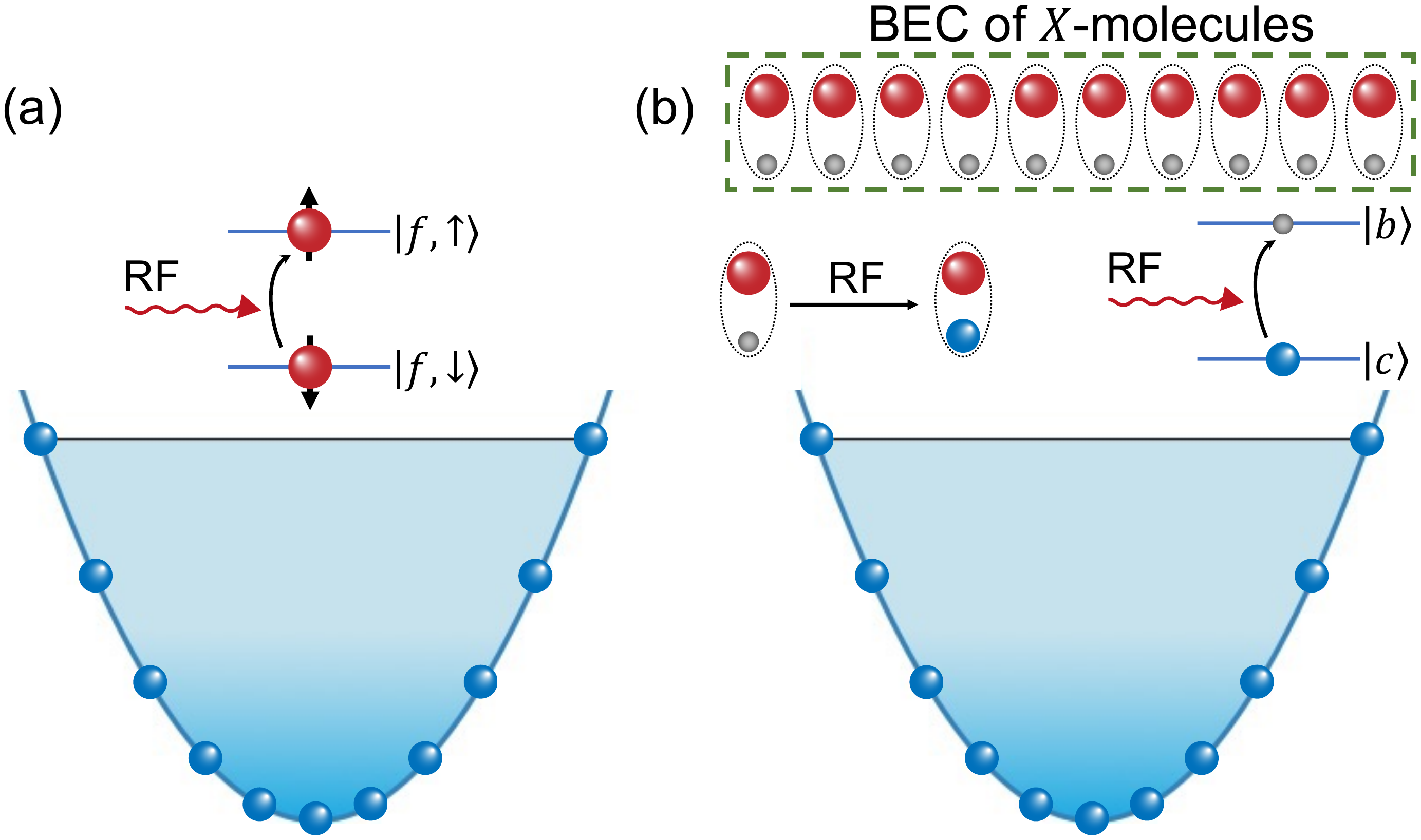}
\caption{ Possible cold-atom setups.  
In the platform of panel (a), we assume that the impurity atom (red ball) has two hyperfine states $\ket{f,\uparrow}$ and $\ket{f,\downarrow}$. Initially, the impurity is prepared in  $\ket{f,\downarrow}$, which is not interacting with the fermionic bath (represented with blue balls). By applying a weak RF-pulse, one drives a transition into $\ket{f,\uparrow}$, which interacts strongly with the Fermi sea. This protocol gives access to the
polaronic spectral properties. The setup of panel (b) is different. Here we
consider the initial state to be a Fermi sea of $c$-fermions and a BEC of $X$%
-molecules that are composed of atoms $f$ and $b$ (grey balls). Now, we assume that $
\ket{c}$ and $\ket{b}$ are two hyperfine states, which are then coupled by a
weak RF-pulse. Such an RF pulse breaks a ``grey-red'' $X$-molecule and leaves behind one strongly interacting ``red'' $f$-fermion and an additional ``blue'' majority $c$-fermion. These ``red'' and additional ``blue'' fermions are created as a pair with a wave function set by the wave function of the original $X$-molecule. As such, this protocol gives access to the molecular spectral properties. }
\label{schematic}
\end{figure}

Having established the reliability of the non-Gaussian approach to the
ground-state properties of 2D Fermi polaron, we move on to discuss dynamics.
We remark that accurate analysis of out-of-equilibrium properties represents
one of the main advantages of our method compared to, for instance, DMC.
Here, we first discuss possible cold-atom experiments that enable one to
measure polaronic and molecular spectral functions, in particular, to probe
the residues $\mathcal{Z}$ and $\mathcal{Z}_B$. We then analyze these
polaronic and molecular spectral properties separately in the following
subsections.

\subsection{Cold-atom platforms}

We begin by discussing the conventional experimental protocol for measuring
polaronic spectral properties [Fig.~\ref{schematic}(a)]. We  will assume that the impurity atom has two hyperfine states, one of which $\ket{f,\downarrow}$ is not coupled to the $c$-fermions, while the other $\ket{f,\uparrow}$ strongly interacts with the bath. The system is initially prepared in $\ket{f,\downarrow}\otimes\ket{\rm FS}$, which is then driven into $|\Psi _{0}\rangle = \ket{f,\uparrow}\otimes\ket{\rm FS}$ by a weak RF-pulse. Then, Ramsey interferometry enables one to probe the dynamical
overlap function $S(t)=e^{i E_\mathrm{FS} t}\langle \Psi
_{0}|\exp (-iHt)|\Psi _{0}\rangle$, where $E_\mathrm{FS}$ is
the total energy of the Fermi sea. 
The impurity spectral
function $\mathcal{A}(\omega)$, also accessible with RF-spectroscopy, is
given by:
\begin{equation}
\mathcal{A}(\omega)=\frac{1}{\pi}\mathrm{Re}\int_{0}^{\infty} dt\,
e^{i\omega t} \, S(t).
\end{equation}
We remark that while we discuss here the setup to probe correlations when
the total momentum is zero $\mathbf{K}_0 = \mathbf{0}$, it can be extended
to explore $\mathbf{K}_0 \neq \mathbf{0}$ (see Ref.~\cite{Dolgirev2020} for
a related discussion).

The protocol to measure molecular properties is different [Fig.~\ref%
{schematic}(b)]. We now choose the initial state to be a Fermi sea of $c$%
-atoms and a BEC of $X$-molecules, composed of $\ket{f}$ and $\ket{b}$
atoms. In contrast to the previous setup, here $\ket{b}$ and $\ket{c}$ are
assumed to be two hyperfine states. As we demonstrate in Appendix~\ref%
{Appendix: Molecular RF spectroscopy} by performing adiabatic elimination of
$b$-fermions, a weak RF-pulse is then described via:
\begin{equation}
H_{\mathrm{RF}}=\frac{\Omega_{\mathrm{RF}} }{L}\sum_{|\mathbf{k}| \lesssim a_X^{-1}}\frac{%
e^{-i\omega t}}{E_{X}-\varepsilon _{\mathrm{imp},\mathbf{k}}-\varepsilon _{%
\mathrm{b},\mathbf{k}}}c_{\mathbf{k}}^{\dagger }f_{-\mathbf{k}}^{\dagger }+%
\mathrm{H.c.}
\end{equation}%
Here $E_X = -1/(2m_{\mathrm{r}}a_X^{2})$ is the binding energy of an $X$%
-molecule; $a_X$ is the corresponding scattering length assumed to be much
smaller than the Fermi wavelength $a_X k_F \ll 1$. The effective coupling $%
\Omega_{\mathrm{RF}}$ is proportional to $\sqrt{N_{X}}$ and the intensity of
the pulse, with $N_{X}$ being the total number of $X$-molecules. One can
view such an RF-pulse as if it substitutes a $b$-atom in a tightly-bound $X$%
-molecule with a $c$-atom. The corresponding dynamical overlap is given by: $%
S_{B}(t)=e^{i (E_\mathrm{FS}+E_F) t}\langle \Psi _{B}|\exp
(-iHt)|\Psi _{B}\rangle$, where the state $|\Psi _{B}\rangle $ has been
defined in the previous section, cf. Eq.~(\ref{B-type molecular wavefunction}%
). The molecular spectral function $\mathcal{A}_B(\omega)$ is then defined
as:
\begin{align}
\mathcal{A}_B(\omega) =\frac{1}{\pi}\text{Re}\int_{0}^{\infty }dt\,
e^{i\omega t} \, S_B(t).
\end{align}
Having introduced all the relevant dynamical quantities, we turn to explore
them in the next subsections.

\begin{figure}[t!]
\includegraphics[width=1\linewidth]{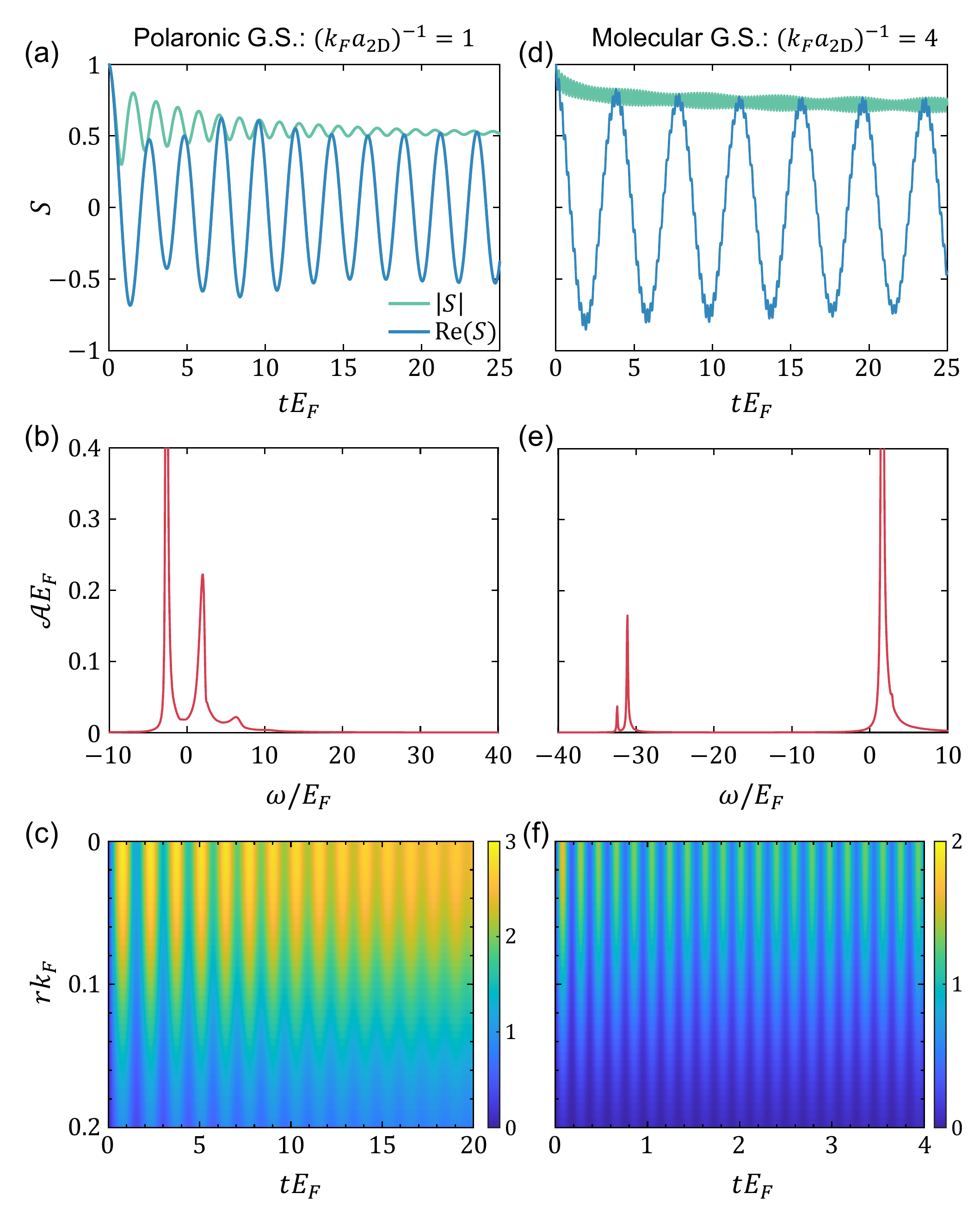}
\caption{ Polaronic spectral properties. Left and right panels correspond to
the cases with the ground state being polaronic $(k_F a_{\mathrm{2D}})^{-1}
= 1$ and with it being molecular $(k_F a_{\mathrm{2D}})^{-1} = 4$,
respectively. Panels (c) and (f) display the evolution of the real-space
density $\mathbf{\protect\delta }\protect\rho (r,t)=\protect\rho (r,t)-%
\protect\rho (r,0)$ (this quantity has the unit of $k_F^2$). Parameters
used: $m_{\mathrm{imp}}=m_{\mathrm{b}}$, $k_{\Lambda}=20k_F$, $\protect\delta%
_p=\left.k_F\right/40$, and $n_{\Lambda }=8$.}
\label{fig 5}
\end{figure}

\subsection{Polaronic spectral properties}

In the co-moving LLP frame, the dynamical overlap reads: $S(t)=\langle
\mathrm{FS}|\bar{\Psi}_{\pmb{0}}(t)\rangle $, where $|\bar{\Psi}_{\pmb{0}}(t)\rangle
=\exp (-iH_{\mathrm{LLP}}t)|\mathrm{FS}\rangle$ -- this latter state is
obtained via the real-time equations of motion detailed in Sec.~\ref{sec:
formalism}. Within the non-Gaussian states, $S(t)$ is computed through $\theta$ and $U$ as:
\begin{equation}
S(t)=e^{-i(\theta(t)-E_\mathrm{FS}t)} \det\left\{\mathrm{I}
_N-[\mathrm{I}_N-U(t)] \Gamma_\mathrm{FS}^{\mathrm{T}}\right\}.
\end{equation}

We begin by considering $(k_{F}a_{\mathrm{2D}})^{-1}=1$, in which case the
ground state is polaronic. Figure~\ref{fig 5}(a) shows the dynamics of Re$%
\,S(t)$ and $|S(t)|$ that display damped oscillatory behavior. At long
times, $|S(t)|$ approaches a finite value, which is nothing but the
quasiparticle residue $\mathcal{Z}$. The spectral function $\mathcal{A}%
(\omega)$ is shown in Fig.~\ref{fig 5}(b). We find that the ground-state
energy, as determined by the position of the sharp peak in $\mathcal{A}%
(\omega)$ (attractive polaron), and the corresponding oscillator strength
are in agreement with the energy and quasiparticle residue calculations in
Figs.~\ref{fig 2} and~\ref{fig 3}. The additional hump in $\mathcal{A}%
(\omega)$ for $\omega > 0$ (repulsive polaron) occurs due to the fact that
the initial state has a finite overlap with the continuum of particle-hole
excitations -- the position and width of this hump determine the frequency
and decay rate of $|S(t)|$ at initial times. Figure~\ref{fig 5}(c) shows the
dynamics of the real-space fermionic density. Upon the abrupt creation of
the impurity at $t = 0$, the density near the impurity initially exhibits
profound oscillations but then reaches a steady state at longer times.

We turn to consider the case with $(k_{F}a_{\mathrm{2D}})^{-1}=4$,
characterized by the ground state being molecular. To properly account for
the finite-momentum nature of the molecular state, here we follow the
previous section and consider $N_f$ + 1 fermions. Figure~\ref{fig 5}(e)
shows the spectral function $\mathcal{A}(\omega)$, which displays two sharp
peaks at $\omega \sim 1.57 E_F$ and $\omega \sim -31.1E_F$ corresponding to
the energies of the repulsive and attractive polarons, respectively, in agreement with the results of Ref.~\cite{schmidt2012}. There
is an additional tiny peak at $\omega \sim -32.5E_F$, which emerges due to
the finite size effects and vanishes in the thermodynamic limit. As we
discuss in the following subsection, this tiny peak corresponds to the
energy of the molecular state. The time-dependent overlap function $S(t)$
exhibits long-lived oscillations shown in Fig.~\ref{fig 5}(d), which can be
understood as arising from quantum beatings between the attractive and
repulsive polarons. We note that the initial state $|\Psi _{0}(t=0)\rangle =|%
\mathrm{FS}\rangle $ has finite overlaps with both of these polaron
branches. As $|\Psi _{0}(t)\rangle $ propagates in time, parts of the wave
function corresponding to the two polarons evolve with different energies,
and since the system seems to never relax to the molecular ground state
locally [Fig.~\ref{fig 5}(d)], it results in long-lived oscillations of the
fermionic density near the impurity, as illustrated in Fig.~\ref{fig 5}(f).

\begin{figure}[t!]
\includegraphics[width=1\linewidth]{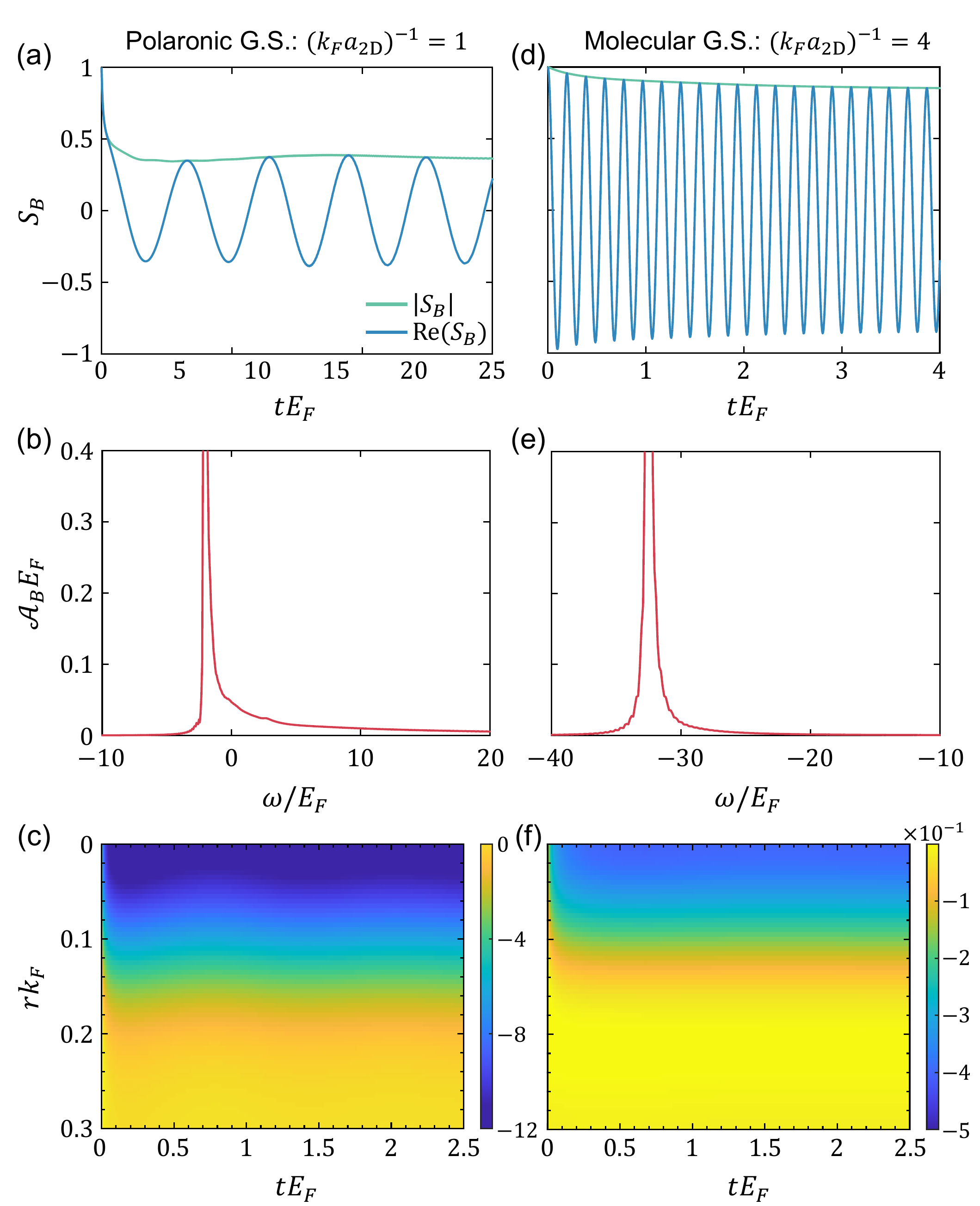}
\caption{ The same as in Fig.~\protect\ref{fig 5}, except here we show
molecular spectral properties. Here we fix $(k_F a_X)^{ -1} = 4$.}
\label{fig 6}
\end{figure}

\subsection{Molecular spectral properties}

Here we do a similar analysis as in the preceding subsection, but now
consider the experimental protocol in Fig.~\ref{schematic}(b) that enables
one to probe molecular spectral properties. 
In the numerical simulations, we consider the initial state $%
|\Psi_{B}(t=0)\rangle =\sum_{\left\vert \mathbf{k}\right\vert >k_{F}}\varphi _{%
\mathbf{k}}^{B}c_{\mathbf{k}}^{\dagger }f_{-\mathbf{k}}^{\dagger }\left\vert
\mathrm{FS}\right\rangle$ in the lab frame, which
describes a deeply bound molecule created on top of the Fermi sea. The corresponding dynamical
overlap function in the co-moving frame becomes: $S_{B}(t)=$ $\langle \bar{%
\Psi}_{B}(t=0)|\bar{\Psi}_{B}(t)\rangle $, where $|\bar{\Psi}_{B}(t)\rangle
=\exp (-iH_{\mathrm{LLP}}t)|\bar{\Psi}_{B}(t=0)\rangle $, and $|\bar{\Psi}%
_{B}(t=0)\rangle =\sum_{\left\vert \mathbf{k}\right\vert >k_{F}}\varphi _{%
\mathbf{k}}^{B}c_{\mathbf{k}}^{\dagger }\left\vert \mathrm{FS}\right\rangle$ is the initial molecule state in the LLP frame, i.e., Eq.~\eqref{eq: psibar B}. Hereafter, for concreteness, we focus on the case $(k_{F} a_{X})^{-1}=4$. The dynamical overlap function $S_{B}(t)$ is calculated
analytically:%
\begin{equation}
S_{B}(t)=e^{-i( \theta (t)-(E_{\mathrm{FS}}+E_{F})t) }\det
\left\{ \mathrm{I}_{N}-[\mathrm{I}_{N}-U(t)]\Gamma _{B}^{\mathrm{T}}\right\}
.
\end{equation}%
Implicit in the discussion below is that the bath is composed of $N_{f}+1$
fermions.

We first discuss the polaronic regime $(k_{F}a_{\mathrm{2D}})^{-1}=1$ -- the
results of our simulations are summarized in Fig.~\ref{fig 6} (left panels).
We find that the frequency of the sharp peak in the molecular spectral
function $\mathcal{A}_{B}(\omega )$ [Fig.~\ref{fig 6}(b)] is $\omega \sim
-2.07E_{F}$, in good agreement with the ground-state energy $%
E_{\mathrm{GS}}\sim -2.09E_{F}$ of $N_{f}+1$ fermions in the total momentum
sector $\mathbf{K}_{0}=\mathbf{0}$. Our results suggest that the
initial tight bound state $|\bar{\Psi}_{B}(t=0)\rangle $ has a finite
overlap {$\sim 0.3$} with the ground state -- this is indicated by the steady-state value of $|S_{B}(t)|$ [Fig.~\ref{fig 6}(a)]. The long tail in $\mathcal{A}_{B}(\omega )$ also implies that the state $|\bar{\Psi}_{B}(t=0)\rangle $ has a substantial spectral weight associated with the continuum of particle-hole excitations of the Fermi sea.

Deep in the molecular phase, $%
(k_{F}a_{\mathrm{2D}})^{-1}=4$ [right panels in Fig.~\ref{fig 6}], the
initial two-body bound state on top of the the undisturbed Fermi sea quickly
relaxes to the true molecular ground state. The sharp peak in the spectral
function $\mathcal{A}_{B}(\omega )$ [Fig.~\ref{fig 6}(e)] is at $\omega \sim
-32.5E_{F}$, which is exactly the molecular ground-state energy. At long
times, $|S_{B}(t)|\sim 0.8$ [Fig.~\ref{fig 6}(d)] -- this value agrees with
the analysis in Fig.~\ref{fig 3}(b) of the molecular residues.

Finally, in both regimes, we find that the density profiles [Fig.~\ref{fig 6}%
(c) and~\ref{fig 6}(f)] show rapid relaxational dynamics.

\section{\label{sec: summary}Summary and outlook}

In this paper, we analyzed both the ground-state and dynamical properties of
Fermi polarons in two spatial dimensions using a new family of non-Gaussian
variational wave functions. We showed that this class of states captures the
polaron-to-molecule transition that emerges as one increases the attractive
interaction strength. Energies of both polaronic and molecular states, as
well as the transition point, are in good agreement with the known
Monte-Carlo simulations. Our theory, in contrast to conventional numerical
methods, enables efficient computation of the polaronic spectral functions,
accessible with RF spectroscopy. In addition to the commonly discussed
quasiparticle spectral function and residue, we introduced molecular
spectral function and residue that help characterize better the nature of
the molecular state. We discussed how these molecular properties could be
measured with RF-like experiments, where we proposed the initial state to
contain a BEC of tightly-bound molecules.

While the analysis in our paper focused on systems of ultracold atoms, we
expect that our results will be relevant for exciton-electron mixtures in
TMD materials. In particular, we anticipate that the proposed experimental
protocol for the molecular spectral properties can be realized in bilayer
TMDs. Indeed, interlayer excitons are relatively long-lived and can be used
to achieve BEC states. Terahertz pulses can then be used to convert these
interlayer excitons into intralayer ones, demonstrate the existence of
Feshbach resonances, and probe the molecular spectral function~\cite{kuhlenkamp2022,tang2021tuning}.

\begin{acknowledgments}
We thank M. Zvonarev, A. Salvador, A. Imamoglu, A. M\"{u}ller, K. Seetharam, I. Esterlis, C. Robens, M. Zwierlein, and R. Schmidt for stimulating discussions, M. M. Parish and J. Levinsen for sharing the data in Ref.~\cite{Parish2013}, and J. Ryckebusch and K. V. Houcke for sharing the data in Ref.~\cite{DMC2D}. T. S. is supported by National Key Research and Development Program of China (Grant No. 2017YFA0718304), by the NSFC (Grants No. 11974363, No. 12135018, and No. 12047503). P. D. and E. D. acknowledge support from the ARO grant number W911NF-20-1-0163 and Harvard/MIT CUA.
\end{acknowledgments}

\appendix

\section{Effective RF-Hamiltonian for the molecular spectral function \label%
{Appendix: Molecular RF spectroscopy}}

Following the setup in Fig.~\ref{schematic}(b) of the main text, here we
derive an effective Hamiltonian that describes the corresponding RF-pulse.
In the rotating frame, the system's Hamiltonian is given by $H=H_{0}+V$,
where
\begin{align}
H_{0}=&\sum_{\mathbf{k}}\left[\left(\varepsilon _{\mathrm{{b},\mathbf{k}}}
-\omega-\Delta \right)c_{\mathbf{k}}^{\dagger }c_{\mathbf{k}}+\tilde{%
\varepsilon}_{\mathbf{k}}b_{\mathbf{k}}^{\dagger }b_{\mathbf{k}}+\varepsilon
_{\mathrm{{imp},\mathbf{k}}} f_{\mathbf{k}}^{\dagger }f_{\mathbf{k}}\right]
\nonumber \\
& + E_XX_{\mathbf{0}}^{\dagger }X_{\mathbf{0}},
\end{align}
and
\begin{align}V =& \sum_{\mathbf{k}}\left(g_{\mathbf{k}} X_{\mathbf{0}}f_{-\mathbf{k}}^{\dagger }b_{%
\mathbf{k}}^{\dagger }+\text{H.c.}\right) + \lambda \sum_{\mathbf{k}}\left(
c_{\mathbf{k}}^{\dagger }b_{\mathbf{k}}+\text{H.c.}\right).
\label{eqn_B_RF}
\end{align}
Here $\omega$ is the frequency of the RF-pulse, $b_{\mathbf{k}}$ ($b_{%
\mathbf{k}}^\dagger$) is the annihilation (creation) operator of the fermion
$b$ with momentum $\mathbf{k}$ and dispersion $\tilde{\varepsilon}_{\mathbf{k%
}} = \varepsilon _{\mathrm{{b},\mathbf{k}}}$, and $\Delta$ is the energy
difference between the states $b$ and $c$. For the BEC of $X$-molecules, we
consider the mode $X_{\mathbf{0}}$ with zero total momentum $\mathbf{{k}={0}}
$ only; it has binding energy $E_X$. The first term in Eq.~\eqref{eqn_B_RF}
describes that an $X$-molecule can recombine into a pair of atoms $f$ and $b$  with the transition amplitude $g_\mathbf{k}$, and vice versa. The second term in Eq.~\eqref{eqn_B_RF} encodes the
RF-pulse that couples the states $\ket{b}$ and $\ket{c}$.

We now use the Schrieffer-Wolf transformation \cite{Schrieffer1966} to
eliminate $V$ and obtain the effective Hamiltonian $H^{\prime }=H_{0}+[V,S]/2$, where the generating operator%
\begin{align}
S =& -\sum_{\mathbf{k}}\frac{1}{\varepsilon _{\mathrm{imp},\mathbf{k}}+%
\tilde{\varepsilon}_{\mathbf{k}}-E_X}\left( g_{\mathbf{k}} X_{\mathbf{0}}f_{-\mathbf{k}%
}^{\dagger }b_{\mathbf{k}}^{\dagger }-\text{H.c.}\right)  \nonumber \\
&-\sum_{\mathbf{k}}\frac{\lambda }{\varepsilon _{\mathrm{b},\mathbf{k}}-%
\tilde{\varepsilon}_{\mathbf{k}}-\omega -\Delta}\left( c_{\mathbf{k}%
}^{\dagger }b_{\mathbf{k}}-\text{H.c.}\right)
\end{align}
satisfies $[H_{0},S]+V=0$. In the explicit form, the effective interaction $%
V_\mathrm{eff}=H^{\prime } -H_{0}$ reads%
\begin{align}
V_\mathrm{eff} =&\frac{1}{2}\sum_{\mathbf{k}}\frac{\lambda \left\langle X_{%
\mathbf{0}}\right\rangle }{\varepsilon _{\mathrm{{imp},\mathbf{k}}}+\tilde{%
\varepsilon}_{\mathbf{k}}-E_X}\left( g_{\mathbf{k}} c_{\mathbf{k}}^{\dagger }f_{-\mathbf{k}%
}^{\dagger }+\text{H.c.}\right)  \nonumber \\
& +\frac{1}{2}\sum_{\mathbf{k}}\frac{\lambda \left\langle X_{\mathbf{0}%
}\right\rangle }{\varepsilon _{\mathrm{b},\mathbf{k}}-\tilde{\varepsilon}_{%
\mathbf{k}}-\omega - \Delta}\left(g_{\mathbf{k}} c_{\mathbf{k}}^{\dagger }f_{-\mathbf{k}%
}^{\dagger }+\text{ H.c.}\right)  \nonumber \\
& +\sum_{\mathbf{k}}\frac{|g_{\mathbf{k}}|^{2}\left\langle X_{\mathbf{0}}\right\rangle ^{2}%
}{\varepsilon _{\mathrm{{imp},\mathbf{k}}}+\tilde{\varepsilon}_{\mathbf{k}%
}-E_X}\left( f_{-\mathbf{k}}^{\dagger }f_{-\mathbf{k}}+b_{\mathbf{k}%
}^{\dagger }b_{\mathbf{k}}-1\right)  \nonumber \\
& +\sum_{\mathbf{k}}\frac{\lambda ^{2}}{\varepsilon _{\mathrm{b},\mathbf{k}}-%
\tilde{\varepsilon}_{\mathbf{k}}-\omega -\Delta}\left( c_{\mathbf{k}%
}^{\dagger }c_{\mathbf{k}}-b_{\mathbf{k}}^{\dagger }b_{\mathbf{k}}\right),
\label{Veff in Schrieffer-Wolf transofrmation}
\end{align}
where the operator $X_{\mathbf{0}}$ is replaced by its expectation value $%
\left\langle X_{\mathbf{0}}\right\rangle$ set by the BEC of $X$-molecules.
Among the four terms appearing in Eq.~\eqref{Veff in Schrieffer-Wolf
transofrmation}, the last two are unimportant as they give small
modifications to the dispersion relations. The second term is expected to be
negligible, if one assumes $|\Delta|\gg|\omega|,|E_{X}|$, while the first
one describes the RF-perturbation. In the original frame, this
latter contribution becomes:%
\begin{figure}[t]
\centering
\includegraphics[width=1\linewidth]{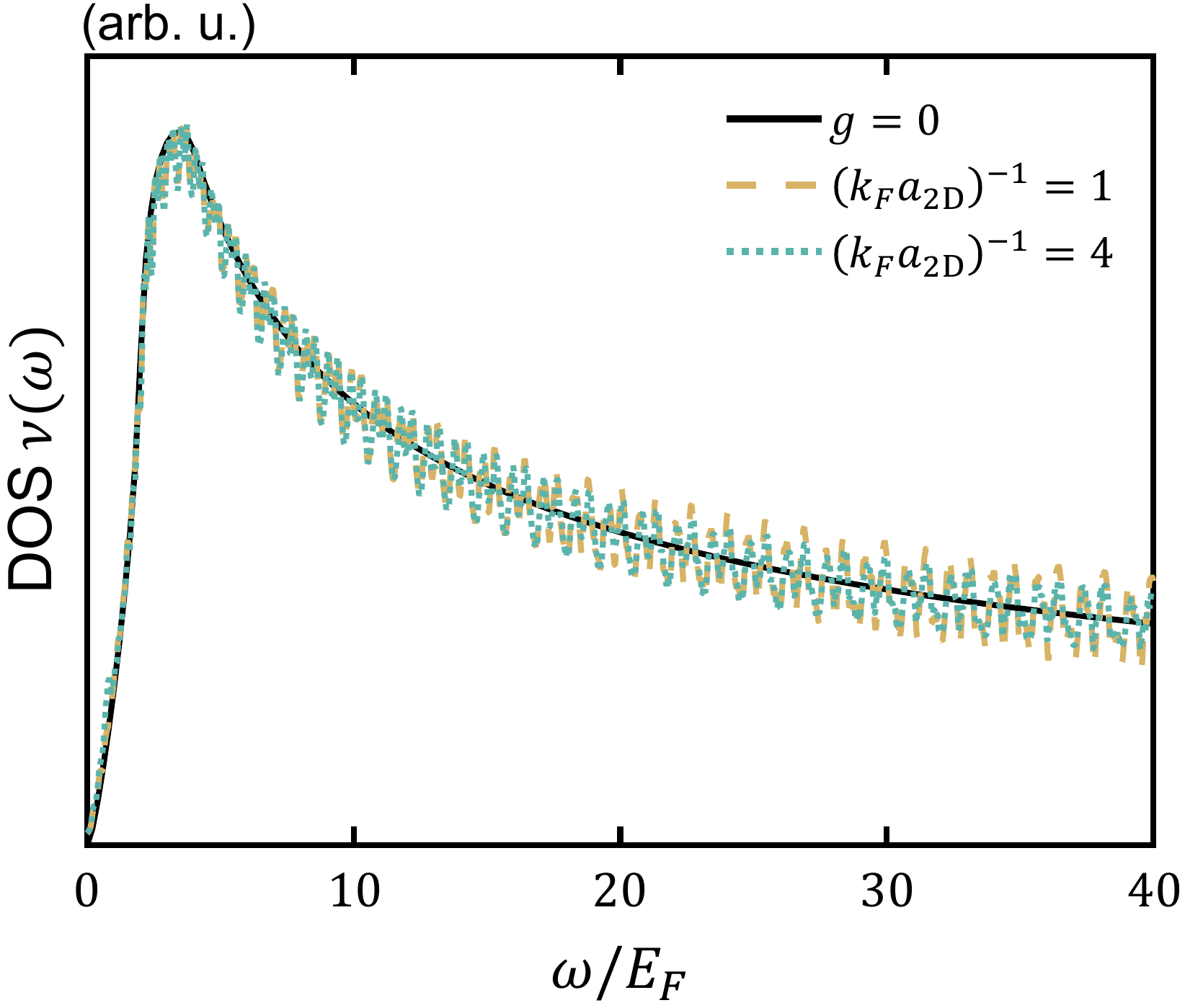}
\caption{Density of states $\protect\nu(\protect\omega)$ (DOS) of collective
modes for various interaction strengths. Parameters used: $\mathrm{K}_0 = 0$%
, $k_{F}=10\protect\delta_p$, $k_{\Lambda }=10k_{F}$, and $n_{\Lambda }=10$.}
\label{fig:fluctuation spectrum}
\end{figure}
\begin{equation}
H_{\mathrm{RF}}=\frac{1}{2}\sum_{\mathbf{k}}\frac{\lambda \left\langle X_{%
\mathbf{0}}\right\rangle }{\varepsilon _{\mathrm{{imp},\mathbf{k}}}+\tilde{%
\varepsilon}_{\mathbf{k}}-E_X}\left( e^{-i\omega t} g_{\mathbf{k}} c_{\mathbf{k}}^{\dagger
}f_{-\mathbf{k}}^{\dagger }+\text{ H.c.}\right) .
\end{equation}
The transition amplitude $g_\mathbf{k}$ can be approximated as a constant, provided that it varies slowly in the range $|\mathbf{k}| \lesssim a_X^{-1}$.

\section{DOS of collective modes}\label{appendix:collective modes}
Collective modes in the 1D Fermi polaron proved particularly important
because they help understand not
only equilibrium but also
far-from-equilibrium phenomena, such as the quantum flutter~\cite%
{Dolgirev2020}. Motivated by this and following Sec.~\ref{subsec_CM_analysis}%
, here we compute the density of states (DOS) $\nu (\omega )=\sum_{n}\delta
(\omega -\omega _{n})$ of collective modes in the 2D polaron problem ($%
\omega_n$ labels energies of the excitations). Figure~\ref{fig:fluctuation
spectrum} shows the result of such an analysis.

We do not find any
particular features in $\nu(\omega)$, such as a sharp peak that could
resemble the quantum flutter. What is surprising is that DOSs for notably
different interaction strengths look similar, which could be because we
restrict our analysis to the momentum sector $\mathbf{K}_{0}=\mathbf{0}$. It
could also be that far-from-equilibrium dynamics in the 2D Fermi polaron are
very different from the 1D case. We leave the analysis for generic total
momenta $\mathbf{K}_{0}\neq \mathbf{0}$ to future work.

\bibliographystyle{apsrev4-2.bst}
\bibliography{bib}

\end{document}